\documentclass[11pt,a4paper]{article}

\usepackage{jcappub,amsfonts,amsmath,amssymb,bm,graphicx,subfigure}

\title{Neutrino spin oscillations in a magnetized Polish doughnut}

\author{Maxim Dvornikov}
\affiliation{Pushkov Institute of Terrestrial Magnetism, Ionosphere and Radiowave Propagation (IZMIRAN), 108840 Moscow, Troitsk, Russia}
\emailAdd{maxdvo@izmiran.ru}

\abstract{
We study the gravitational scattering of ultrarelativistic neutrinos
off a rotating supermassive black hole (BH) surrounded by a thick
magnetized accretion disk. Neutrinos interact electroweakly with background
matter and with the magnetic field in the disk since neutrinos are
supposed to possess nonzero magnetic moments. The interaction with
external fields results in neutrino spin oscillations. We find that
the toroidal magnetic field, inherent in the magnetized Polish doughnut,
does not cause a significant spin-flip for any reasonable strengths
of the toroidal component. The reduction of the observed neutrino flux,
owing to neutrino spin oscillations, is predicted. A poloidal component
of the magnetic field gives the main contribution to the modification
of the observed flux. The neutrino interaction with matter, rotating
with relativistic velocities, also changes the flux of neutrinos.
We briefly discuss the idea of the neutrino tomography of magnetic
field distributions in accretion disks near BHs.
}


\arxivnumber{2307.10126}

\begin{document}

\maketitle

\section{Introduction\label{sec:INTR}}

The experimental achievements, e.g., in refs.~\cite{Fuk98,Ahm02}, confirmed
that neutrinos are massive particles and there is a nonzero mixing
between different neutrino generations. These results open a window
to explore physics beyond the standard model. The nonzero neutrino
masses inevitably result in nontrivial neutrino electromagnetic properties.
Namely, it is shown in refs.~\cite{LeeScr77,DvoStu04} that, even in the minimally
extended standard model supplied with a right handed neutrino, there
is a small neutrino magnetic moment. It means that a neutrino is no
longer considered as a purely electrically neutral particle. For example,
a nonzero neutrino magnetic moment leads to the particle spin precession
in an external magnetic field.

The change of the neutrino polarization in external fields has dramatic
consequences for the observability of these particles. In frames of
the standard model, a neutrino is created as a left-handed particle,
i.e. its spin is opposite to the neutrino momentum. If the neutrino
polarization changes, the particle becomes right-handed, i.e. sterile.
Hence, we shall observe the effective reduction of the initial neutrino
flux. This process is called neutrino spin oscillations. Spin oscillations
of astrophysical neutrinos were thoroughly studied by many authors
(see, e.g., ref.~\cite{Raf96}). For example,
the upper bound on the neutrino magnetic moment by studying spin oscillations
of supernova (SN) neutrinos was obtained in ref.~\cite{KuzMikOkr09}.

It was demonstrated in ref.~\cite{VolVysOku86} that, besides the interaction with
a magnetic field, neutrino spin oscillations can be affected by the
neutrino electroweak interaction with background matter. The gravitational
interaction, although it is quite weak, can also cause the precession
of a fermion spin. It was shown in ref.~\cite{Wal72} that the motion of a spinning
body in a curved spacetime deviates from geodesic. Assuming that a
fermion is an elementary particle, it was obtained in ref.~\cite{PomKhr98} that
its spin is parallel transported along the particle trajectory.
The quantum theory of the fermion spin in curved spacetime was developed in ref.~\cite{ObuSilTer17}.

Using the quasiclassical approach in ref.~\cite{PomKhr98}, neutrino spin
oscillations in curved spacetime were studied in ref.~\cite{Dvo06} within
the General Relativity (GR), and its extensions in refs.~\cite{AlaNod15,Cha15,MasLam21,Pan22}.
In this work, we continue the studies in refs.~\cite{Dvo23a,Dvo23b,Dvo23c},
where spin effects in the neutrino gravitational scattering off a
rotating black hole (BH) are accounted for. In a gravitational scattering,
both incoming and outgoing neutrinos are in the asymptotically flat
spacetime. Therefore, their spin states are well defined.

As shown in refs.~\cite{Dvo23a,Dvo23b,Dvo23c}, the magnetic field in an
accretion disk gives the main contribution to neutrino spin oscillations.
However, we used quite simple models of accretion disk in those papers.
It should be noted that an accretion disk should be thick in order
that both magnetic field and background matter are able to influence
neutrino spin oscillations.

In the present work, we rely on the thick accretion disk model call
a Polish doughnut~\cite{AbrJarSik78}. The
magnetized version of a Polish doughnut was developed in ref.~\cite{Kom06}. We also account for the possibility
of the presence of a poloidal magnetic field inside the accretion
disk. One of the models for such a field was proposed in ref.~\cite{FraMei09}.
Other models of accretion disks are reviewed in ref.~\cite{AbrFra13}.

The present work was motivated by the recent observations of the BH
shadows in the centers of M87 and our Galaxy in refs.~\cite{Aki19,Aki22}.
It is the first direct test of GR in the strong field limit. It was suggested in ref.~\cite{BerGin81}
that, besides photons emitted by an accretion disk, it can be a source
of high energy neutrinos. It should be noted that there are active
searches for high energy neutrinos emitted in active galactic nuclei,
e.g., in ref.~\cite{Abb22}.

The image of such a disk, observed in a neutrino telescope, should
account for both strong gravitational lensing of particles, as well
as the neutrino spin precession in external fields, which converts
active left neutrinos to sterile ones. Thus, in a neutrino telescope,
we shall observe a different picture compared to an optical image
plotted, e.g., in ref.~\cite{ArmRey03}. Another possibility to probe
spin effects in the neutrino gravitational scattering is to observe
the lensing of SN neutrinos by a supermassive BH (SMBH) in the center
of our Galaxy. For example, such a
possibility was discussed in ref.~\cite{MenMocQui06}.

Our work is organized as follows. We recall how to describe the trajectory
of ultrarelativistic particles scattered off a rotating BH in section~\ref{sec:NUSCATT}.
In section~\ref{sec:NUSPINEVOL}, we represent the neutrino spin evolution
in external fields in curved spacetime. The structure of external
fields in the accretion disk in given in section~\ref{sec:DISK}. We
fix the characteristics of the external fields and a neutrino in section~\ref{sec:PARAM}.
In section~\ref{sec:RES}, we present the results of numerical simulations.
Finally, we conclude in section~\ref{sec:CONCL}. The main expressions
for a magnetized Polish doughnut are listed in appendix~\ref{sec:MAGPOLDOU}.
In appendix~\ref{sec:BELOW}, we show how to use the symmetry of the
system to reconstruct the spin precession of some neutrinos.

\section{Neutrino scattering off a rotating BH\label{sec:NUSCATT}}

In this section, we briefly remind how to describe the motion of ultrarelativistic
neutrinos scattered off a rotating BH.

The spacetime of a rotating BH has the Kerr metric which is written down in the following
form using the Boyer-Lindquist coordinates $x^{\mu}=(t,r,\theta,\phi)$:
\begin{equation}\label{eq:Kerrmetr}
  \mathrm{d}s^{2}=g_{\mu\nu}\mathrm{d}x^{\mu}\mathrm{d}x^{\nu}=
  \left(
    1-\frac{rr_{g}}{\Sigma}
  \right)\mathrm{d}t^{2}+
  2\frac{rr_{g}a\sin^{2}\theta}{\Sigma}\mathrm{d}t\mathrm{d}\phi-
  \frac{\Sigma}{\Delta}\mathrm{d}r^{2}-\Sigma\mathrm{d}\theta^{2}-  
  \frac{\Xi}{\Sigma}\sin^{2}\theta\mathrm{d}\phi^{2},
\end{equation}
where
\begin{equation}\label{eq:ingmet}
  \Delta=r^{2}-rr_{g}+a^{2},
  \quad
  \Sigma=r^{2}+a^{2}\cos^{2}\theta,
  \quad
  \Xi=
  \left(
    r^{2}+a^{2}
  \right)
  \Sigma+rr_{g}a^{2}\sin^{2}\theta.
\end{equation}
Here $r_{g}$ is the Schwarzschild radius. The mass of BH is $M=r_{g}/2$
and its spin, which is along the $z$-axis, is $J=Ma$, where $0<a<M$.

The motion of a test ultrarelativistic particle in the metric in eq.~(\ref{eq:Kerrmetr})
can be found in quadratures~\cite{Cha83}. It has three integrals
of motion: the particle energy, $E$, its angular momentum, $L$,
and the Carter constant, $Q$. In the scattering problem, $Q>0$.
It was shown in ref.~\cite{GraLupStr18} that form of the trajectory can be inferred
from the integral expressions,
\begin{align}
  z & \int\frac{\mathrm{d}x}{\pm\sqrt{R(x)}}=
  \int\frac{\mathrm{d}t}{\pm\sqrt{\Theta(t)}},
  \label{eq:rt}
  \\
  \phi & =z\int\frac{(x-zy)\mathrm{d}x}{\pm\sqrt{R(x)}(x^{2}-x+z^{2})}+
  \frac{y}{z}\int\frac{\mathrm{d}t}{\pm\sqrt{\Theta(t)}(1-t^{2})}
  \label{eq:phirt}
\end{align}
where
\begin{align}\label{eq:RTh}
  R(x) & =
  \left(
    x^{2}+z^{2}-yz
  \right)^{2}-(x^{2}-x+z^{2})
  \left[
    w+
    \left(
      z-y
    \right)^{2}
  \right],
  \nonumber
  \\
  \Theta(t) & =(t_{-}^{2}+t^{2})(t_{+}^{2}-t^{2}),
  \nonumber
  \\
  t_{\pm}^{2} & =\frac{1}{2z^{2}}
  \left[
    \sqrt{(z^{2}-y^{2}-w)^{2}+4z^{2}w}\pm(z^{2}-y^{2}-w)
  \right],
\end{align}
We use the dimensionless variables in eqs.~(\ref{eq:rt})-(\ref{eq:RTh}):
$x=r/r_{g}$, $y=L/r_{g}E$, $z=a/r_{g}$, $w=Q/r_{g}^{2}E^{2}$,
and $t=\cos\theta$.

The strategy for finding the neutrino trajectory is the following.
First, one computes the $x$-integral in eq.~(\ref{eq:rt}) numerically.
Then, $\theta=\arccos t$ is obtained from eq.~(\ref{eq:rt}) using
the Jacobi elliptic functions. At this stage, we should account for
inversions of the trajectory in the equatorial plane. Finally, eq.~(\ref{eq:phirt})
is used to get $\phi$. The details of calculations are provided in ref.~\cite{Gru14}.

\section{Neutrino spin evolution in curved spacetime under the influence of
external fields\label{sec:NUSPINEVOL}}

In this section, we consider the description of the neutrino polarization
when a particle scatters off a rotating BH surrounded by a realistic
thick magnetized accretion disk.

The covariant four vector of the spin $S^{\mu}$ of a neutrino, which
interacts with an electromagnetic field and background matter, obeys
the following equation in curved spacetime~\cite{Dvo13},
\begin{align}\label{eq:BMTcurvedst}
  \frac{\mathrm{D}S^{\mu}}{\mathrm{d}\tau}= & 2\mu
  \left(
    F^{\mu\nu}S_{\nu}-U^{\mu}U_{\nu}F^{\nu\lambda}S_{\lambda}
  \right)+
  \sqrt{2}G_{\mathrm{F}}E^{\mu\nu\lambda\rho}
  G_{\nu}U_{\lambda}S_{\rho},
\end{align}
where $\mathrm{D}S^{\mu}=\mathrm{d}S^{\mu}+\Gamma_{\alpha\beta}^{\mu}S^{\alpha}\mathrm{d}x^{\beta}$
is the covariant differential, $\Gamma_{\alpha\beta}^{\mu}$ are the
Christoffel symbols, $U^{\mu}=\tfrac{\mathrm{d}x^{\mu}}{\mathrm{d}\tau}$
is the neutrino four velocity in the world coordinates, $\tau$ is the proper
time, $E^{\mu\nu\lambda\rho}=\tfrac{1}{\sqrt{-g}}\varepsilon^{\mu\nu\lambda\rho}$
is the covariant antisymmetric tensor in a curved spacetime, $g=\det(g_{\mu\nu})$
is the determinant of the metric tensor, $F_{\mu\nu}$ is the tensor
of an external electromagnetic field, $\mu$ is the neutrino magnetic
moment, $G_{\mathrm{F}}=1.17\times10^{-5}\,\text{GeV}^{-2}$ is the
Fermi constant, and $G_{\mu}$ is the covariant effective potential
of the neutrino electroweak interaction with a background matter.
Equation~(\ref{eq:BMTcurvedst}) is valid for both massive and massless
(ultrarelativistic) neutrinos.

The neutrino polarization is defined in the locally Minkowskian frame
$x_{a}=e_{a}^{\,\mu}x_{\mu}$. The vierbein vectors $e_{a}^{\,\mu}$
satisfy the relation, $\eta_{ab}=e_{a}^{\,\mu}e_{b}^{\,\nu}g_{\mu\nu}$,
where $\eta_{ab}=(1,-1,-1,-1)$ is the Minkowski metric tensor. One
can check that $e_{a}^{\,\mu}$ have the form,
\begin{align}\label{eq:vierbKerr}
  e_{0}^{\ \mu}= &
  \left(
    \sqrt{\frac{\Xi}{\Sigma\Delta}},0,0,
    \frac{arr_{g}}{\sqrt{\Delta\Sigma\Xi}}
  \right),
  \quad
  e_{1}^{\ \mu}=
  \left(
    0,\sqrt{\frac{\Delta}{\Sigma}},0,0
  \right),
  \nonumber
  \\
  e_{2}^{\ \mu}= &
  \left(
    0,0,\frac{1}{\sqrt{\Sigma}},0
  \right),
  \quad
  e_{3}^{\ \mu}=
  \left(
    0,0,0,\frac{1}{\sin\theta}\sqrt{\frac{\Sigma}{\Xi}}
  \right),
\end{align}
where $\Delta$, $\Sigma$, and $\Xi$ are given in eq.~(\ref{eq:ingmet}).

We rewrite eq.~(\ref{eq:BMTcurvedst}) in this Minkowskian frame
making the boost to the particle rest frame, where the invariant three
vector of the neutrino polarization, $\bm{\zeta}$, is defined,
\begin{equation}\label{eq:nuspinrot}
  \frac{\mathrm{d}\bm{\bm{\bm{\zeta}}}}{\mathrm{d}t}=
  2(\bm{\bm{\bm{\zeta}}}\times\bm{\bm{\bm{\Omega}}}).
\end{equation}
The vector $\bm{\bm{\Omega}}=\bm{\bm{\Omega}}_{g}+\bm{\bm{\Omega}}_{\mathrm{em}}+\bm{\bm{\Omega}}_{\mathrm{matt}}$,
which incorporates the neutrino interaction with external fields including
gravity, has the following form for an ultrarelativistic neutrino:
\begin{align}\label{eq:vectG}
  \bm{\bm{\Omega}}_{g} & = \frac{1}{2U^{t}}
  \left[
    \mathbf{b}_{g}+\frac{1}{1+u^{0}}
    \left(
      \mathbf{e}_{g}\times\mathbf{u}
    \right)
  \right],
  \nonumber
  \\
  \bm{\bm{\Omega}}_{\mathrm{em}} & =\frac{\mu}{U^{t}}
  \left[
    u^{0}\mathbf{b}-
    \frac{\mathbf{u}(\mathbf{u}\mathbf{b})}{1+u^{0}}+
    (\mathbf{e}\times\mathbf{u})
  \right],
  \nonumber
  \\
  \bm{\bm{\Omega}}_{\mathrm{matt}} & =
  \frac{G_{\mathrm{F}}\mathbf{u}}{\sqrt{2}U^{t}}
  \left(
    g^{0}-\frac{(\mathbf{gu})}{1+u^{0}}
  \right),
\end{align}
where $(\mathbf{e}_{g},\mathbf{b}_{g})=G_{ab}=\gamma_{abc}u^{c}$,
$\gamma_{abc}=\eta_{ad}e_{\,\mu;\nu}^{d}e_{b}^{\,\mu}e_{c}^{\,\nu}$
are the Ricci rotation coefficients, the semicolon stays for the covariant
derivative, $(\mathbf{e},\mathbf{b})=f_{ab}=e_{a}^{\ \mu}e_{b}^{\ \nu}F_{\mu\nu}$
is the electromagnetic field tensor in the locally Minkowskian frame,
$(u^{0},\mathbf{u})=u^{a}=e_{\,\mu}^{a}U^{\mu}$, and $(g^{0},\mathbf{g})=g^{a}=e_{\,\mu}^{a}G^{\mu}$.

The electric and magnetic fields are $e_i = f_{0i}$ and $b_i = - \varepsilon_{ijk} f_{jk}$, where $\varepsilon_{ijk}$ is the antisymmetric tensor in the flat spacetime. The explicit form of these vectors depends on the original configuration of the electromagnetic field in world coordinates $F_{\mu\nu}(x^\mu)$. These vectors for certain models of electromagnetic fields in an accretion disk versus $r$ and $\theta$ are given shortly in section~\ref{sec:DISK}; cf. eqs.~\eqref{eq:bztor}-\eqref{eq:ebsoph}.

Instead of solving the precession eq.~(\ref{eq:nuspinrot}), we deal
with the effective Schr\"{o}dinger equation to describe the neutrino polarization,
\begin{equation}\label{eq:Schreq}
  \mathrm{i}\frac{\mathrm{d}\psi}{\mathrm{d}x}=
  \hat{H}_{x}\psi,
  \quad
  \hat{H}_{x}=-\mathcal{U}_{2}
  (\bm{\bm{\sigma}}\cdot\bm{\bm{\Omega}}_{x})
  \mathcal{U}_{2}^{\dagger},
\end{equation}
where $\bm{\bm{\sigma}}=(\sigma_{1},\sigma_{2},\sigma_{3})$ are the
Pauli matrices, $\bm{\bm{\Omega}}_{x}=r_{g}\bm{\bm{\Omega}}\tfrac{\mathrm{d}t}{\mathrm{d}r}$,
and $\mathcal{U}_{2}=\exp(\mathrm{i}\pi\sigma_{2}/4)$. Equation~(\ref{eq:Schreq})
is rewritten in dimensionless variables and adapted for the scattering
problem. It is convenient to rewrite the vector $\bm{\bm{\Omega}}_{x}$
in the form,
\begin{align}\label{eq:Omegax}
  \bm{\bm{\Omega}}_{x}^{(g)} & =\frac{1}{2}
  \left[
    \tilde{\mathbf{b}}_{g}+(\tilde{\mathbf{e}}_{g}\times\mathbf{v})
  \right],
  \nonumber
  \\
  \bm{\bm{\Omega}}_{x}^{(\mathrm{em})} & =V_{\mathrm{B}}
  \left[
    l^{0}\mathbf{b}-\mathbf{v}(\mathbf{lb})+(\mathbf{e}\times\mathbf{l})
  \right],
  \nonumber
  \\
  \bm{\bm{\Omega}}_{x}^{(\mathrm{matt})} & =V_{m}\mathbf{l}       
  \left[
    g^{0}-(\mathbf{gv})
  \right],
\end{align}
where the vectors $(l^{0},\mathbf{l})=l^{a}=\frac{\mathrm{d}t}{\mathrm{d}r}\frac{u^{a}}{U^{t}}$,
$\mathbf{v}=\frac{\mathbf{u}}{1+u^{0}}$, $\tilde{\mathbf{e}}_{g}=\mathbf{e}_{g}\frac{r_{g}}{U^{t}}\tfrac{\mathrm{d}t}{\mathrm{d}r}$
and $\tilde{\mathbf{b}}_{g}=\mathbf{b}_{g}\frac{r_{g}}{U^{t}}\tfrac{\mathrm{d}t}{\mathrm{d}r}$,
are finite for an ultrarelativistic neutrino. Note that $\tilde{\mathbf{e}}_{g}$
and $\tilde{\mathbf{b}}_{g}$ are the linear functions of $\tfrac{\mathrm{d}x^{\mu}}{\mathrm{d}r}$
which can be calculated using the results of section~\ref{sec:NUSCATT}. The explicit
form of $l^{a}$ and $\mathbf{v}$ is provided in ref.~\cite{Dvo23a}.

The scalar quantity $V_{\mathrm{B}}$ depends on the configuration
of the magnetic field in the disk, which is discussed shortly in section~\ref{sec:DISK}.
The parameter $V_{m}=\tfrac{G_{\mathrm{F}}\rho}{\sqrt{2}m_{p}r_{g}}$
for a hydrogen plasma, where $\rho$ is the mass density of the disk
and $m_{p}$ is the proton mass.

The effective spinor $\psi$ in eq.~(\ref{eq:Schreq}) has form $\psi_{-\infty}^{\mathrm{T}}=(1,0)$
for incoming neutrinos. Such a spinor corresponds to a left polarized
active neutrino. For a scattered particle, after solving eq.~(\ref{eq:Schreq})
along the neutrino trajectory, it becomes $\psi_{+\infty}^{\mathrm{T}}=(\psi_{+\infty}^{(\mathrm{R})},\psi_{+\infty}^{(\mathrm{L})})$.
The survival probability, i.e. the probability that a neutrino remains
left polarized in the wake of the scattering, is $P_{\mathrm{LL}}=|\psi_{+\infty}^{(\mathrm{L})}|^{2}$. 

\section{External fields in an accretion disk\label{sec:DISK}}

In this section, we discuss the properties of the background matter
and the magnetic fields in an accretion disk which a neutrino interacts
with.

We treat electroweak interaction of a neutrino with background fermions
in the forward scattering approximation. We mention in section~\ref{sec:NUSPINEVOL}
that this interaction is characterized by the four potential $G^{\mu}$,
which has the following form in the hydrogen plasma:
\begin{equation}\label{eq:Gmu}
  G^{\mu}=\sum_{f=e,p}q_{f}J_{f}^{\mu},
\end{equation}
where $J_{f}^{\mu}=n_{f}U_{f}^{\mu}$ are the hydrodynamic currents,
$n_{f}$ are the invariant fermions densities, $U_{f}^{\mu}$ are
their four velocities in the disk, and $q_{f}$ are the constants
which are found in the explicit form in ref.~\cite{DvoStu02}. We suppose
in eq.~(\ref{eq:Gmu}) that matter is unpolarized. We take that $n_{e}=n_{p}$
because of the plasma electroneutrality and $U_{e}^{\mu}=U_{p}^{\mu}$,
i.e. there is no differential rotation between the components of plasma.

In the model of a Polish doughnut, one has that $U_{f}^{\mu}=(U_{f}^{t},0,0,U_{f}^{\phi})$;
cf. eq.~(\ref{eq:UB}). Thus, using eq.~(\ref{eq:vierbKerr}), we
get that the nonzero components of $g^{a}$ in eq.~(\ref{eq:Omegax})
are
\begin{align}\label{eq:gcomp}
  g^{0} & =\frac{\sqrt{x^{2}+z^{2}\cos^{2}\theta}
  \sqrt{x^{2}-x+z^{2}}U_{f}^{t}}
  {\sqrt{z^{2}\cos^{2}\theta(x^{2}-x+z^{2})+z^{2}x^{2}+z^{2}x+x^{4}}},  
  \nonumber
  \\
  g^{3} & =\frac{\sin\theta
  \left[
    r_{g}U_{f}^{\phi}
    \left(
      z^{2}\cos^{2}\theta(x^{2}-x+z^{2})+z^{2}x^{2}+z^{2}x+x^{4}
    \right)-
    U_{f}^{t}xz
  \right]}
  {\sqrt{x^{2}+z^{2}\cos^{2}\theta}\sqrt{z^{2}\cos^{2}\theta
  (x^{2}-x+z^{2})+z^{2}x^{2}+z^{2}x+x^{4}}}.
\end{align}
The mass density $\rho$, which enters in the coefficient $V_{m}$,
in given in eq.~(\ref{eq:rhopm}).

Now, we discuss the neutrino interaction with magnetic fields. The
toroidal magnetic is inherent in the magnetized Polish doughnut model.
The four vector of such a magnetic field is $B^{\mu}=(B^{t},0,0,B^{\phi})$;
cf. eq.~(\ref{eq:UB}). The electromagnetic field tensor of this
field has the only nonzero components $F^{r\theta}=-F^{\theta r}$. Based on eq.~(\ref{eq:vierbKerr}),
we get that the nonzero component of $\mathbf{b}$ in eq.~(\ref{eq:Omegax})
is
\begin{align}\label{eq:bztor}
  b_{3}= & -\frac{U_{f}^{t}r_{g}^{2}\sqrt{2p_{m}^{(\mathrm{tor})}}}
  {\sin\theta(1-\Omega l_{0})
  \sqrt{(x^{2}+z^{2}\cos^{2}\theta)(x^{2}-x+z^{2})}}
  \nonumber
  \\
  & \times
  \big\{
    \lambda_{0}^{2}(x^{2}-x+z^{2}\cos^{2}\theta)+
    2\lambda_{0}xz\sin^{2}\theta
    \notag
    \\
    & -
    \sin^{2}\theta
    \left[
      (x^{2}+z^{2})(x^{2}+z^{2}\cos^{2}\theta)+xz^{2}\sin^{2}\theta
    \right]
  \big\}^{1/2},
\end{align}
where $\lambda_{0}=l_{0}/r_{g}$, $l_{0}$ is the constant angular
momentum in the disk (see appendix~\ref{sec:MAGPOLDOU}), $\Omega$
is given in eq.~(\ref{eq:Omegadisk}), and $p_{m}^{(\mathrm{tor})}$
is the magnetic pressure of the toroidal field present in eq.~(\ref{eq:rhopm}).
The vector $\mathbf{e}=0$. The dimensionless coefficient $V_{\mathrm{B}}$
in eq.~(\ref{eq:Omegax}) is $V_{\mathrm{B}}=\mu/r_{g}$.

A poloidal magnetic field is not a part of the magnetized Polish doughnut
model. It is, however, known that a superposition of poloidal and
toroidal components makes the resulting magnetic field more stable. That is why
we include a poloidal field in our calculations. We consider two models
for a poloidal field.

First, we take a field which asymptotically tends to a constant one
parallel to the rotation axis of BH at the infinity. The vector potential
for such a field is given in eq.~(\ref{eq:Atphi}). The nonzero components
of the dimensionless vectors $\mathbf{e}$ and $\mathbf{b}$ are
\begin{align}\label{eq:eb}
  e_{1} = & f_{\mathrm{B}}(x)\frac{z
  \left[
    z^{2}\cos^{4}\theta(z^{2}-x^{2})+
    \cos^{2}\theta(z^{4}+2z^{2}x^{2}-3x^{4})-z^{2}x^{2}+x^{4}
  \right]}
  {2\sqrt{z^{2}\cos^{2}\theta(x^{2}-
  x+z^{2})+z^{2}x^{2}+z^{2}x+x^{4}}(x^{2}+z^{2}\cos^{2}\theta)^{2}},
  \nonumber
  \\
  e_{2} = & \frac{f_{\mathrm{B}}(x)xz^{3}\sin2\theta
  \sqrt{x^{2}-x+z^{2}}(1+\cos^{2}\theta)}
  {2\sqrt{z^{2}\cos^{2}\theta(x^{2}-x+z^{2})+z^{2}x^{2}+z^{2}x+x^{4}}
  (x^{2}+z^{2}\cos^{2}\theta)^{2}},
  \nonumber
  \\
  b_{1} = & \frac{f_{\mathrm{B}}(x)\cos\theta
  \left[
    (x^{2}+z^{2}\cos^{2}\theta)^{2}(x^{2}-x+z^{2})+x(x^{4}-z^{4})
  \right]}
  {\sqrt{z^{2}\cos^{2}\theta(x^{2}-x+z^{2})+z^{2}x^{2}+z^{2}x+x^{4}}
  (x^{2}+z^{2}\cos^{2}\theta)^{2}},
  \nonumber
  \\
  b_{2} = & f_{\mathrm{B}}(x)\sin\theta\sqrt{x^{2}-x+z^{2}}
  \notag
  \\
  & \times
  \frac{
  \left[
    z^{4}\cos^{4}\theta(1-2x)+z^{2}\cos^{2}\theta(z^{2}-x^{2}-4x^{3})-  
    z^{2}x^{2}-2x^{5}
  \right]}
  {2\sqrt{z^{2}\cos^{2}\theta(x^{2}-x+z^{2})+z^{2}x^{2}+z^{2}x+x^{4}}
  (x^{2}+z^{2}\cos^{2}\theta)^{2}},
\end{align}
where, following ref.~\cite{BlaPay82}, we introduce the additional factor
$f_{\mathrm{B}}(x)=x^{-5/4}$ to provide the scaling of the magnetic
field with the distance $B\propto r^{-5/4}$. The dimensionless coefficient
$V_{\mathrm{B}}$ in eq.~(\ref{eq:Omegax}), corresponding to such
a field, is $V_{\mathrm{B}}=\mu B_{0}r_{g}$, where $B_{0}$ is the
magnetic field strength near BH at $x\sim1$.

Second, we consider a poloidal field generated by the vector potential
in eq.~(\ref{eq:Aphi}). In this case, the nonzero components of
vectors $\mathbf{e}$ and $\mathbf{b}$ in eq.~(\ref{eq:Omegax})
are
\begin{align}\label{eq:ebsoph}
  e_{1}= & -\frac{zxbr_{g}\partial_{r}\rho}
  {(x^{2}+z^{2}\cos^{2}\theta)\sqrt{z^{2}\cos^{2}\theta(x^{2}-x+z^{2})+
  z^{2}x^{2}+z^{2}x+x^{4}}},
  \nonumber
  \\
  e_{2}= & -\frac{zxb\partial_{\theta}\rho}
  {(x^{2}+z^{2}\cos^{2}\theta)\sqrt{x^{2}-x+z^{2}}
  \sqrt{z^{2}\cos^{2}\theta(x^{2}-x+z^{2})+z^{2}x^{2}+z^{2}x+x^{4}}},
  \nonumber
  \\
  b_{1}= & -\frac{b\partial_{\theta}\rho}
  {\sin\theta\sqrt{z^{2}\cos^{2}\theta(x^{2}-x+z^{2})+
  z^{2}x^{2}+z^{2}x+x^{4}}},
  \nonumber
  \\
  b_{2}= & \frac{\sqrt{x^{2}-x+z^{2}}br_{g}\partial_{r}\rho}
  {\sin\theta\sqrt{z^{2}\cos^{2}\theta(x^{2}-x+z^{2})+
  z^{2}x^{2}+z^{2}x+x^{4}}},
\end{align}
where $\rho$ and $b$ are given in eqs.~(\ref{eq:rhopm}) and~(\ref{eq:Aphi}).
The density derivatives with respect to $r$ and $\theta$ can be
calculated analytically using eq.~(\ref{eq:rhopm}). However, the
corresponding expressions are quite cumbersome and, thus, we omit
them. The dimensionless coefficient $V_{\mathrm{B}}$ in eq.~(\ref{eq:Omegax})
is $V_{\mathrm{B}}=\mu/r_{g}$ for this field component.

\section{Parameters of the system\label{sec:PARAM}}

In this section, we specify the values of the parameters of a neutrino
and external fields, as well as describe some details of calculations.

We suppose that a neutrino is a Dirac particle possessing a nonzero
magnetic moment $\mu$. Its value is $\mu=10^{-13}\mu_{\mathrm{B}}$,
where $\mu_{\mathrm{B}}$ is the Bohr magneton. This magnetic moment
is below the best astrophysical upper bounds for neutrino magnetic
moments established in ref.~\cite{Via13}. We assume that neutrinos do
not have transition magnetic moments, i.e. we study spin oscillations
within one neutrino flavor, e.g., for electron neutrinos.

Neutrinos interact with background matter within the standard model.
The effective potential in given in eq.~(\ref{eq:Gmu}). We suppose
that the disk consists of the electroneutral hydrogen plasma. In this
case, the effective potential in eq.~(\ref{eq:Gmu}) depends on the
electron number density only. The maximal density in the disk is taken
as $n_{e}^{(\mathrm{max})}=10^{18}\,\text{cm}^{-3}$. Such a value
is consistent with the observations carried out in ref.~\cite{Jia19}
for SMBH with $M=10^{8}M_{\odot}$.

The matter density and velocity depend on $r$ and $\theta$ in the
Polish doughnut model for the accretion disk. The typical density
distribution is shown in figure~\ref{fig:nepmdistr} for different
spins of BH. The accretion disk model has a free parameter $K$ [see
eq.~(\ref{eq:rhopm})]. We vary $K$ to make $n_{e}^{(\mathrm{max})}=10^{18}\,\text{cm}^{-3}$
for all spins of BH.

The toroidal magnetic field has the same configuration in our calculations
[see eqs.~(\ref{eq:bztor}) or~(\ref{eq:UB})]. We vary the
parameter $K_{m}$ in eq.~(\ref{eq:rhopm}) to reach $|\mathbf{B}|_{\mathrm{max}}^{(\mathrm{tor})}=320\,\text{G}$.
This magnetic field strength is $|\mathbf{B}|_{\mathrm{max}}^{(\mathrm{tor})}\sim10^{-2}B_{\mathrm{Edd}}$,
where
\begin{equation}
  B_{\mathrm{Edd}} = 10^4\thinspace\text{G}\times
  \left(
    \frac{M}{10^9M_\odot}
  \right)^{-1/2},
\end{equation}
is the Eddington limit for the magnetic
field in the vicinity of BH~\cite{Bes10}, which is $B_{\mathrm{Edd}}\approx 3.2\times 10^4\,\text{G}$
for $M=10^{8}M_{\odot}$.

We consider two models of the poloidal magnetic field in the disk.
First, the vector potential is in eq.~(\ref{eq:Atphi}) [see also
eq.~(\ref{eq:eb})]. Second, $A_{\mu}$ is given eq.~(\ref{eq:Aphi})
[see also eq.~(\ref{eq:ebsoph})]. In the latter case, we choose the
parameter $b$ so that $|\mathbf{B}|_{\mathrm{max}}^{(\mathrm{pol})}=320\,\text{G}$.
It means that the toroidal and poloidal fields have comparable strengths.

The initial flux of neutrinos is emitted from the point $(r_{s,}\theta_{s},\phi_{s})=(\infty,\pi/2,0)$.
All incoming neutrinos are left polarized. First, we reconstruct the
neutrino trajectory, namely, the dependence $\theta(x)$ using eq.~(\ref{eq:rt}).
Then, we use the $\theta(x)$-dependence in eqs.~(\ref{eq:gcomp})-(\ref{eq:ebsoph}),
which converts eq.~(\ref{eq:Schreq}) into the ordinary differential
equation. Finally, we integrate eq.~(\ref{eq:Schreq}) with $\bm{\bm{\Omega}}_{x}$
in eq.~(\ref{eq:Omegax}) along each neutrino trajectory using the
two-step Adams--Bashforth developed in ref.~\cite{Dvo23b}. We deal
with $\sim2.5\times10^{3}$ test particles in each run.

\section{Results\label{sec:RES}}

In this section, we present the results of numerical simulations with
the parameters given in section~\ref{sec:PARAM}.

We mention in section~\ref{sec:INTR} that active neutrinos are left-polarized,
i.e. their spin is opposite to the particle momentum. If the neutrino
spin precesses in an external field, a particle becomes sterile, i.e.
we will observe the effective reduction of the neutrino flux. If we
define the survival probability $P_{\mathrm{LL}}$, i.e. the probability
to remain left-handed after scattering, the observed neutrino flux
is $F_{\nu}=P_{\mathrm{LL}}F_{0}$, where $F_{0}$ is the flux of
`scalar' particles, i.e. particles which propagate along geodesics
lines without necessity to track their polarization. Hence, $F_{0}$
is the observed flux at no spin oscillations. The detailed study of $F_{0}$ was carried
out in ref.~\cite{Gru14}. Our main goal is to obtain $F_{\nu}/F_{0}$
and examine its dependence on the parameters of the system.

First, we mention that, in the scattering of ultrarelativistic neutrinos,
the gravitational interaction only does not result in the neutrino
spin-flip. This result was obtained in ref.~\cite{Lam05} in the situation of
a weak gravitational lensing.
This fact was confirmed in refs.~\cite{Dvo23a,Dvo23b,Dvo23c} for a strong gravitational lensing of neutrinos. Since our
current simulations reveal the same feature, we omit the corresponding
plot for $F_{\nu}/F_{0}$, which is trivial.

\begin{figure}
  \centering
  \includegraphics[scale=0.3]{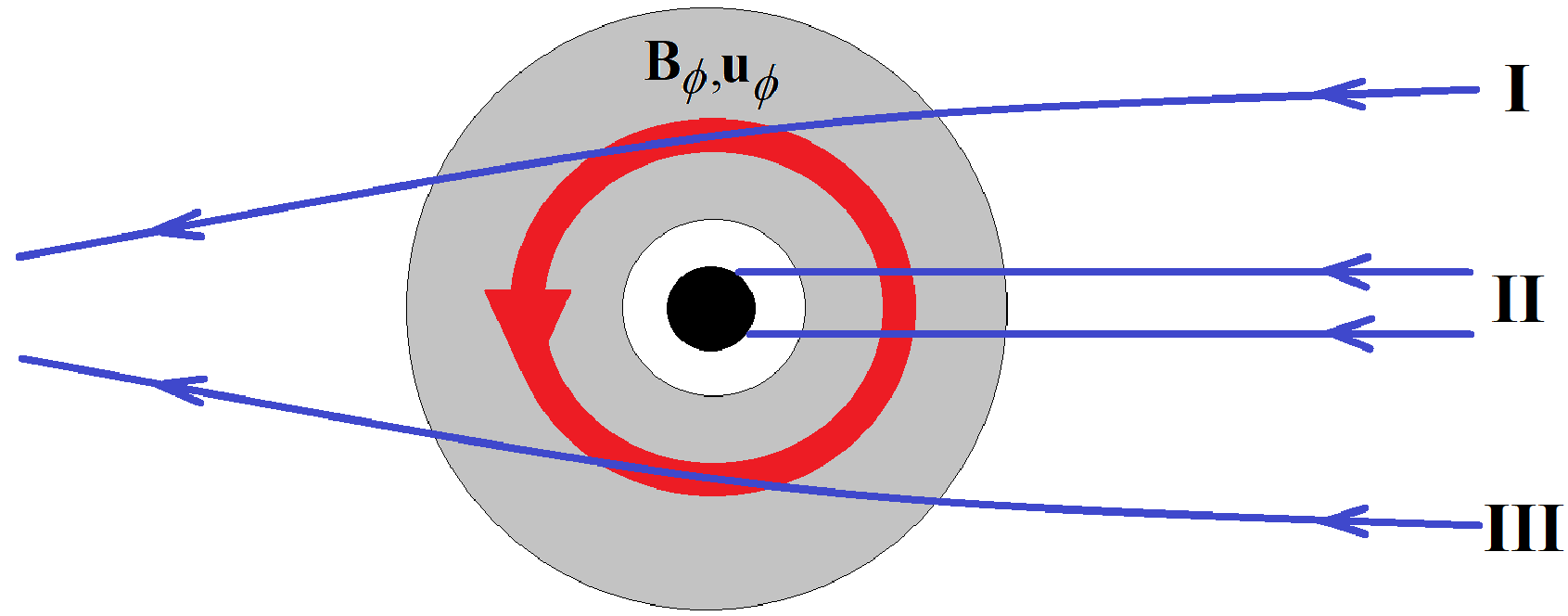}
\caption{Schematic illustration of the neutrino interaction with a magnetized
Polish doughnut, shown with gray color. The toroidal magnetic field
and plasma velocity are concentrated mainly within the thin red torus.
Neutrino trajectoties, depicted with blue color, are separated into
three regions. Regions~I and~III cointain particles which interact
with the longitudinal magnetic field. Neutrinos in region~II interact
with the transverse field, but they mainly fall to BH, shown with
a black blob.\label{fig:schemscatt}}
\end{figure}

The toroidal magnetic field is inherent in a Polish doughnut accretion
disk. The impact of the toroidal magnetic field on neutrino spin oscillations
turns out to be negligible. The process of neutrino scattering and
spin oscillations in a toroidal field is schematically depicted in
figure~\ref{fig:schemscatt}. As seen in figure~\ref{fig:nepmdistr},
the toroidal magnetic field is concentrated in a relatively thin torus.
Spin oscillations are efficient if there is a significant transverse
component of a magnetic field. In figure~\ref{fig:schemscatt}, one
can see that particles in regions~I and~III in the incoming flux
will interact mainly with a longitudinal magnetic field in their scattering.
Despite particles in the region~II interact with a transverse field,
they are mainly within the shadow of BH. Hence, such neutrinos
fall into BH and will not contribute to the observed flux. That is
why the toroidal field with a reasonable strength does not modify
the picture of spin oscillations.

If we consider the neutrino interaction with matter in the combination
with the toroidal field, such external fields do not result in a significant
spin oscillation. Indeed, it was found in ref.~\cite{VolVysOku86} that the interaction
with matter only does not cause spin oscillations. It can only shift
the resonance point. Thus, we omit the plots showing $F_{\nu}/F_{0}$
for the case of the toroidal field and background matter.

\begin{figure}
  \centering
  \subfigure[]
  {\label{fig:f2a}
  \includegraphics[scale=.35]{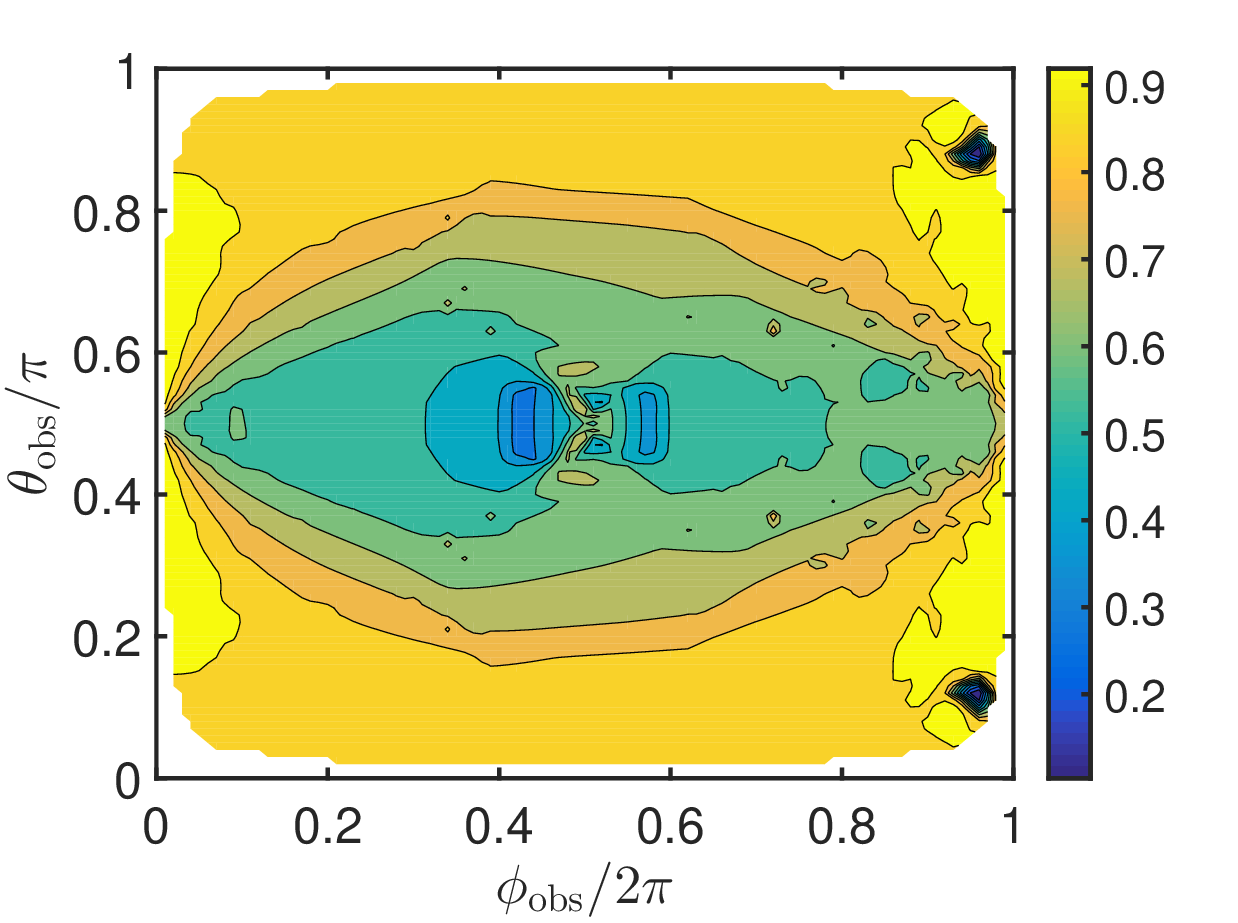}}
  \subfigure[]
  {\label{fig:f2b}
  \includegraphics[scale=.35]{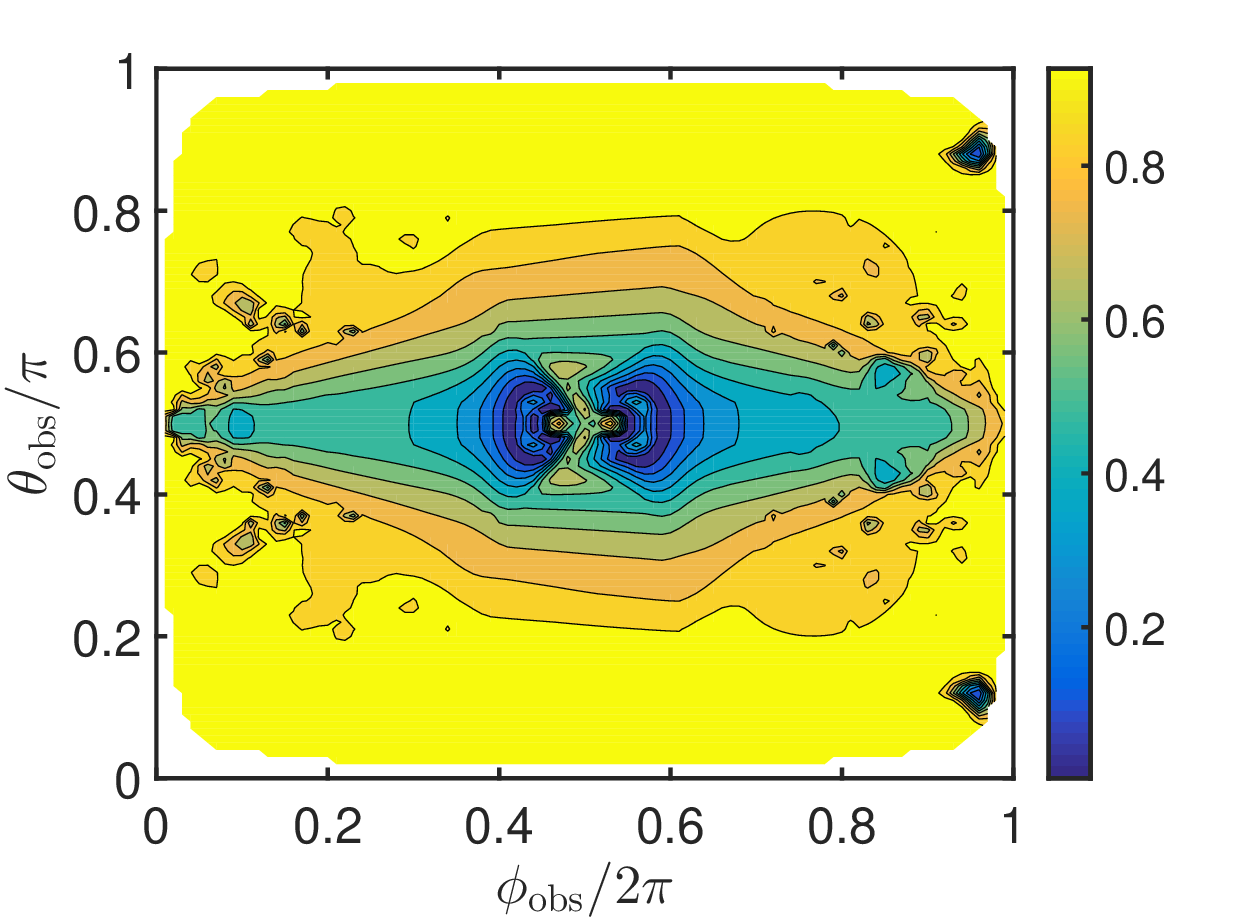}}
  \\
  \subfigure[]
  {\label{fig:f2c}
  \includegraphics[scale=.35]{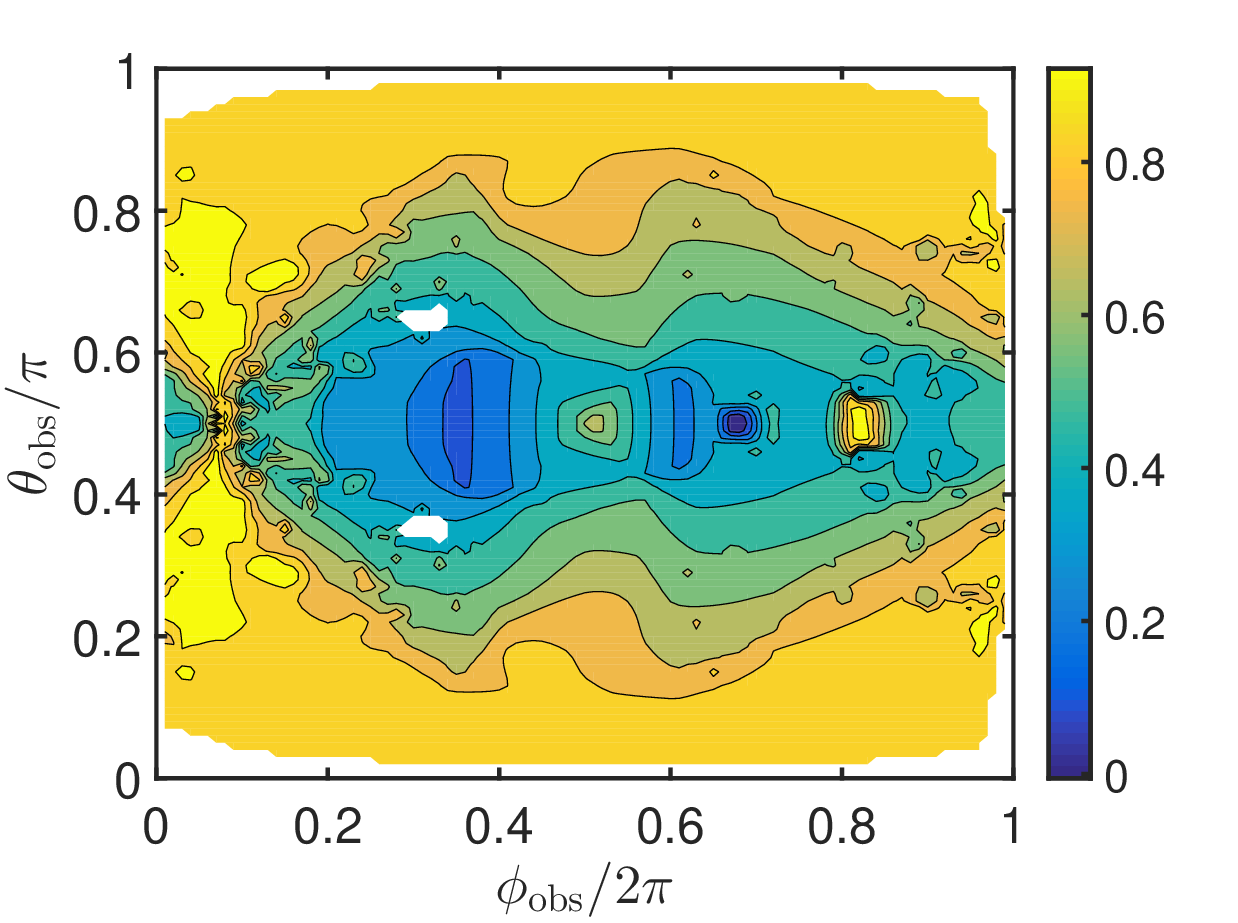}}
  \subfigure[]
  {\label{fig:f2d}
  \includegraphics[scale=.35]{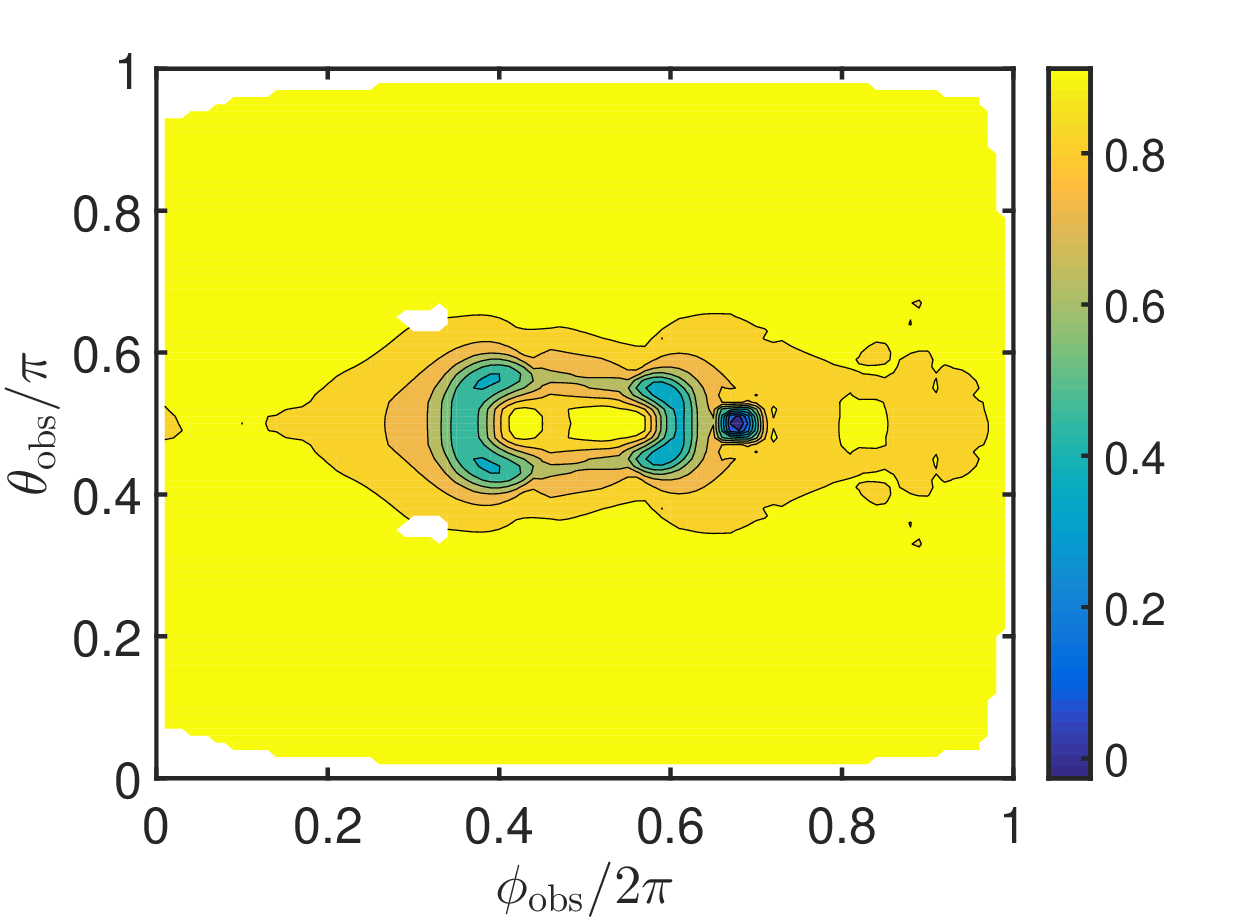}}
  \\
  \subfigure[]
  {\label{fig:f2e}
  \includegraphics[scale=.35]{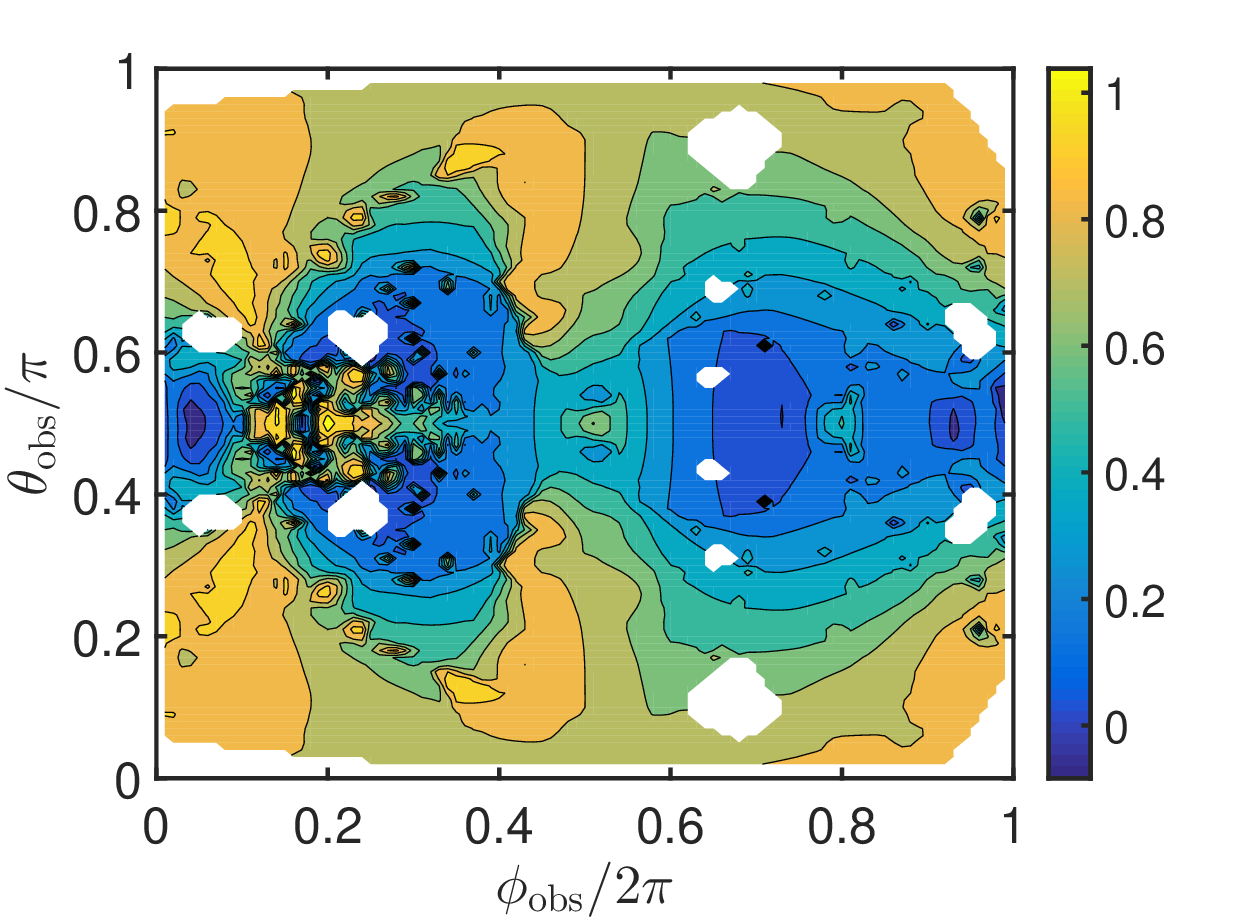}}
  \subfigure[]
  {\label{fig:f2f}
  \includegraphics[scale=.35]{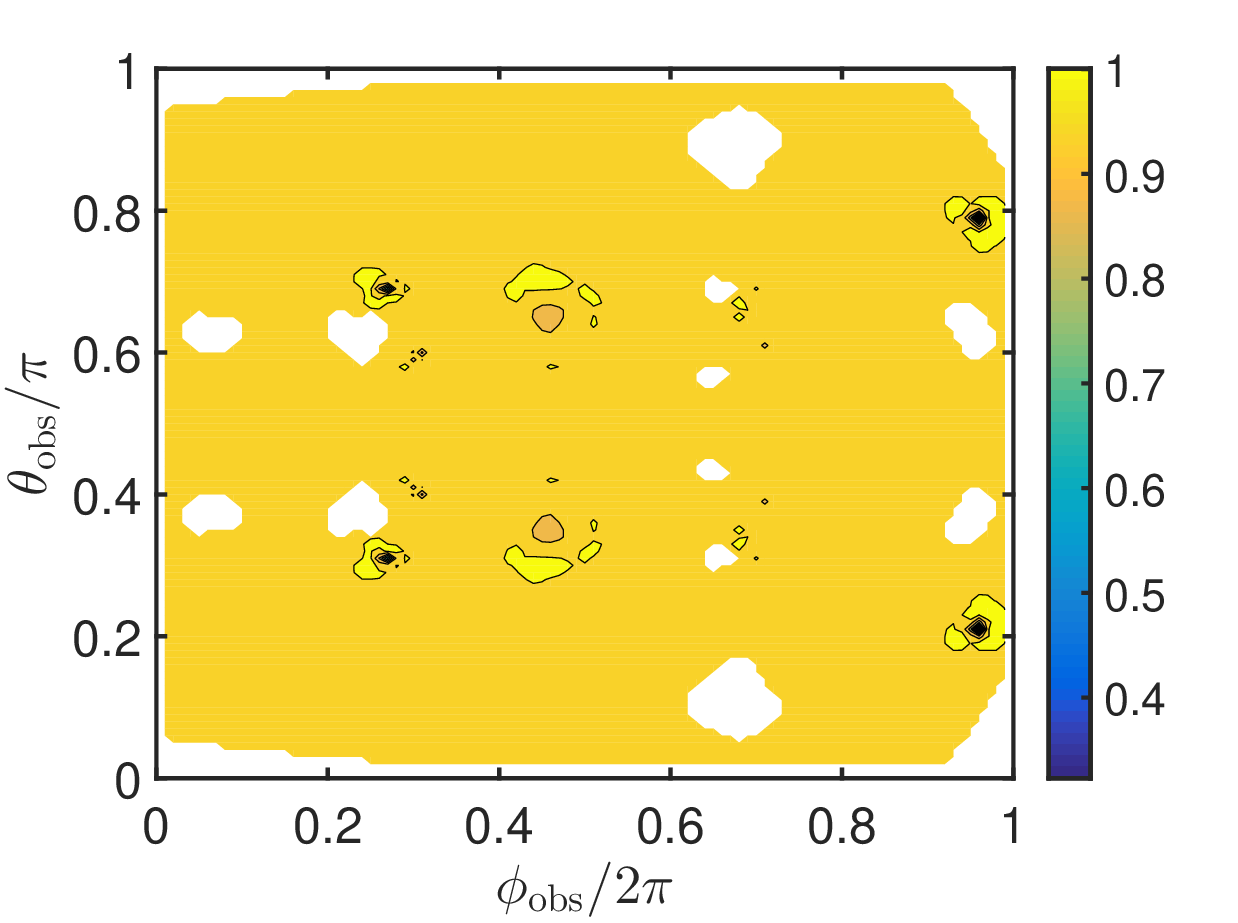}}
  \protect 
\caption{The observed neutrino flux normalized by $F_{0}$ for neutrinos scattered
off a rotating SMBH with $M=10^{8}M_{\odot}$. The maximal number
density in the disk is $10^{18}\,\text{cm}^{-3}$. The maximal strengths
of toroidal and poloidal fields are $320\,\text{G}$. The neutrino
magnetic moment is $10^{-13}\mu_{\mathrm{B}}$. Panels~(a) and~(b):
$a=2\times10^{-2}M$ ($z=10^{-2}$); panels~(c) and~(d): $a=0.5M$
($z=0.25$); panels~(e) and~(f): $a=0.9M$ ($z=0.45$). Panels~(a),
(c), and~(e) correspond to the poloidal field in eq.~(\ref{eq:Atphi});
panels~(b), (d), and~(f) -- in eq.~(\ref{eq:Aphi}). The rest
of the parameters is the same as in figure~\ref{fig:nepmdistr}.\label{fig:fluxpol}}
\end{figure}

The ratio of fluxes for neutrino gravitational scattering and the
interaction with matter under the influence of both toroidal and poloidal
field is shown in figure~\ref{fig:fluxpol}. We depict $F_{\nu}/F_{0}$
for different spins of BH and various models of a poloidal field.
The neutrino fluxes are given versus $\theta_{\mathrm{obs}}=\theta(t\to+\infty)$
and $\phi_{\mathrm{obs}}=\phi(t\to+\infty)$, which are angular coordinates
of an outgoing neutrino. First, we note that the plots in figure~\ref{fig:fluxpol}
are symmetric with respect to the equatorial plane. It is the consequence
of the fact that the flux of incoming neutrinos is parallel to the
equatorial plane (see also appendix~\ref{sec:BELOW}).

Since the incoming neutrinos are emitted from the point $\theta_{s}=\pi/2$
and $\phi_{s}=0$, the image of BH is mainly in the point $\theta_{\mathrm{obs}}\approx\pi/2$
and $\phi_{\mathrm{obs}}\approx\pi$. This feature is seen, especially,
in figures~\ref{fig:f2a} and~\ref{fig:f2b}, which
correspond to an almost nonrotating BH. Despite figures~\ref{fig:f2a}
and~\ref{fig:f2b} correspond to an almost Schwarzschild BH,
these plots are not fully symmetric with respect to the vertical line
$\phi_{\mathrm{obs}}=\pi$. It happens since neutrinos interact with
the rotating accretion disk, where plasma moves with relativistic
velocities. Again, referring to schematic plot in figure~\ref{fig:schemscatt},
it means that particles in regions~I and~III have different diagonal
elements in the effective Hamiltonian $\hat{H}_{x}$ since the velocities
of plasma and neutrinos are in the same (region~I) and the opposite
(region~III) directions.

If we compare figures~\ref{fig:f2a}, \ref{fig:f2c},
and~\ref{fig:f2e} with figures~\ref{fig:f2b}, \ref{fig:f2d},
and~\ref{fig:f2f}, we can see that spin oscillations for
the poloidal field in eq.~(\ref{eq:Atphi}) are more intense than
for eq.~(\ref{eq:Aphi}). For example, spin oscillations are almost
absent in figure~\ref{fig:f2f}. This feature is explained by
the fact that the poloidal field in eq.~(\ref{eq:Aphi}) has a significant
strength only in a small volume in the vicinity of BH (see, e.g.,
figure~\ref{fig:pmplosoph}), whereas the field in eq.~(\ref{eq:Atphi})
slowly decreases towards the outer edge of the disk.

One can also see in figure~\ref{fig:fluxpol} that there is a strong
dependence of $F_{\nu}/F_{0}$ on the spin of BH, especially for the
poloidal field in eq.~(\ref{eq:Aphi}). It happens since the greater
$a$ is, the closer $|\mathbf{B}|_{\mathrm{max}}^{(\mathrm{pol})}$
to the inner radius of the accretion disk is; cf. figure~\ref{fig:pmplosoph}.

There are small white areas in figure~\ref{fig:fluxpol} which increase
with $a$. They appear because the 2D interpolation cannot correctly process
plots with local insufficient number of points. This shortcoming can
be eliminated by a significant enhancement of the test particles number.

\section{Conclusion\label{sec:CONCL}}

We have studied neutrino scattering off a rotating SMBH surrounded
by a thick magnetized accretion disk. Neutrinos were supposed to interact
with BH gravitationally, as well as with matter of the disk electroweakly.
The nonzero magnetic moment of Dirac neutrinos also allowed them the
interact with the magnetic field in the disk. The interaction with
external fields resulted in the precession of the neutrino spin which,
in its turn, led to the effective reduction of the observed neutrino
flux.

Our main goal was to find the observed neutrino flux accounting for
spin oscillations, $F_{\nu}$. It is shown in figure~\ref{fig:fluxpol}
for different spins of BH and various configurations of magnetic fields
in the disk. We normalize $F_{\nu}$ to $F_{0}$, which is the flux
of scalar particles, i.e. at no spin oscillations.

In the present work, we have used the model of the accretion disk
called the magnetized Polish doughnut~\cite{Kom06}. This model predicts
the self consistent distributions of the matter density, angular velocity,
and the toroidal magnetic field in the disk. Hence, our present results
are the advance compared to the calculations in refs.~\cite{Dvo23a,Dvo23b,Dvo23c}
where the characteristics of accretion disks were taken from different
sources.

Analogously to~\cite{Dvo23a,Dvo23b,Dvo23c}, here, we have confirmed
that only gravitational interaction does not cause the spin-flip of
ultrarelativistic neutrinos in their gravitational scattering. This
fact is valid in a wide range of spins of BH. Thus, the only source
of neutrino spin oscillations is the neutrino interaction with the
magnetic field. We recall that the interaction with matter in case
of ultrarelativistic neutrinos does not cause a spin-flip either.
Moreover, we have found that the toroidal magnetic field within the
magnetized Polish doughnut does not result in a significant change
of the observed neutrino flux. It is the consequence of the quite
compact location of the toroidal field.

A configuration with the only toroidal component is known to be unstable.
That is why we have assumed that a poloidal magnetic field is present
in the disk. We have considered two models of the poloidal field;
cf. eqs.~(\ref{eq:Atphi}) and~(\ref{eq:Aphi}). The typical strengths
of toroidal and poloidal fields were taken to be equal.

It should be noted that the presence of a strong poloidal magnetic field in an accretion disk around SMBHs is required in ref.~\cite{BlaZna77} for the formation of jets from these objects. A nonzero poloidal component in an accretion disk was obtained using the MHD simulations in curved spacetime in ref.~\cite{VilHaw03}. The analytical estimates for the relation between $B_r$ and $B_z$ components of the poloidal field can be also derived (see, e.g., ref.~\cite{BisLov07}).

Large scale magnetic fields in a thin accretion disk were studied in ref.~\cite{CaoSpr13}. However, a significant neutrino spin-flip is unlikely to be caused by extenal fields in a thin accretion disk since the neutrino path inside such a disk is quite short.

Our results in figure~\ref{fig:fluxpol} show that spin effects are
more sizable for the poloidal field in eq.~(\ref{eq:Atphi}). We
have also revealed the dependence of the observed fluxes on the spin
of BH. For example, one can see in figure~\ref{fig:f2f} that
the observed flux is almost unchanged for a rapidly rotating SMBH
with the poloidal field in eq.~(\ref{eq:Aphi}).

It should be noted that, in our simulations, we have used quite moderate
strengths of magnetic fields, as well as the matter density which
is observed near some SMBHs~\cite{Jia19}. The neutrino magnetic
moment was taken to be below the current astrophysical upper bound~\cite{Via13}.
It makes our results quite plausible. 

Comparing figure~\ref{fig:fluxpol} and figure~\ref{fig:nepmdistr},
as well as figure~\ref{fig:pmplosoph}, we can see that neutrino spin
oscillations are the effective tool for the tomography of the distribution
of the magnetic field in the vicinity of BH. The structure of a magnetic
field, observed with help of neutrinos, is seen especially clearly
for a relatively slowly rotating BH; cf. figures~\ref{fig:f2a}
and~\ref{fig:f2b}. Moreover, if one compares the fluxes in
points, which are symmetric with respect to line $\phi_{\mathrm{obs}}=\pi$,
we can extract the information about the accretion disk rotation.

Our results can be useful for the exploration of external fields in
the vicinity of BHs with help of neutrinos using existing or future
neutrino telescopes~\cite{Abe21,Abu22}. The penetrating power of
neutrinos is much higher than that of photons. Perhaps, in future,
neutrino telescopes will make a serious competition with the facilities
like the Event Horizon Telescope in the studies of BHs vicinities.

We can apply our results to constrain the quantity $\mu B_0$ for neutrinos from a core-collapsing SN, which is expected in our galaxy~\cite{RozVisCap21}. Here, we rely on the poloidal magnetic field model in eq.~\eqref{eq:Atphi} for the definitiveness. The predicted $F_\text{pred}(\theta_\text{obs},\phi_\text{obs})$ and observed $F_\text{obs}(\theta_\text{obs},\phi_\text{obs})$ neutrino fluxes in a certain direction $(\theta_\text{obs},\phi_\text{obs})$ are related by $F_\text{obs} = P_\mathrm{LL}F_\text{pred}$, where $P_\mathrm{LL}(\theta_\text{obs},\phi_\text{obs}|\mu B_0)$ is the survival probability shown in figure~\ref{fig:fluxpol}. Using the constraint on the magnetic field in the vicinity of SMBH in Sgr A$^{*}$, $B<10^2\,\text{G}$, obtained in ref.~\cite{Eat13}, as well as figure~\ref{fig:fluxpol}, the upper bound on $\mu$ can be derived. However, it requires $F_\text{obs}$ which will be available only after the observation of SN neutrinos.

\appendix

\section{Magnetized Polish doughnut accounting for the poloidal magnetic field\label{sec:MAGPOLDOU}}

In this appendix, we review the main properties of a magnetized Polish
doughnut. The very detailed description of this model is given in ref.~\cite{Kom06}.
That is why we represent only the major expressions since the signature
of our metric is mainly $(+,-,-,-)$, which is different from ref.~\cite{Kom06}.

All parameters of the disk depend on $r$ and $\theta$ owing to the
axial symmetry of the metric in eq.~(\ref{eq:Kerrmetr}). The electromagnetic
field tensor has the form,
\begin{equation}\label{eq:Fmunudisk}
  F_{\mu\nu}=E_{\mu\nu\alpha\beta}U_{f}^{\alpha}B^{\beta},
\end{equation}
where $E^{\mu\nu\alpha\beta}=\tfrac{\varepsilon^{\mu\nu\alpha\beta}}{\sqrt{-g}}$
is the antisymmetric tensor in curved spacetime with $\varepsilon^{tr\theta\phi}=1$.
The four vectors of the fluid velocity in the disk and the toroidal
magnetic field are $U_{f}^{\mu}=(U_{f}^{t},0,0,U_{f}^{\phi})$ and
$B^{\mu}=(B^{t},0,0,B^{\phi})$. We assume that the specific angular
momentum of a particle in the disk $l=L/E$ is constant, $l=l_{0}$.
It allows one the find the components of $U_{f}^{\mu}$ and $B^{\mu}$,
\begin{align}\label{eq:UB}
  U_{f}^{t} & =\sqrt{
  \left|
    \frac{\mathcal{A}}{\mathcal{L}}
  \right|}
  \frac{1}{1-l_{0}\Omega},
  \quad
  U_{f}^{\phi}=\Omega U_{f}^{t},
  \nonumber
  \\
  B^{\phi} & =\sqrt{\frac{2p_{m}^{(\mathrm{tor})}}{|\mathcal{A}|}},
  \quad
  B^{t}=l_{0}B^{\phi},
\end{align}
where $\mathcal{L}=g_{tt}g_{\phi\phi}-g_{t\phi}^{2}$, $\mathcal{A}=g_{\phi\phi}+2l_{0}g_{t\phi}+l_{0}^{2}g_{tt}$,
and
\begin{equation}\label{eq:Omegadisk}
  \Omega=-\frac{g_{t\phi}+l_{0}g_{tt}}{g_{\phi\phi}+l_{0}g_{t\phi}},
\end{equation}
 is the angular velocity in the disk.

The disk density $\rho$ and the magnetic pressure $p_{m}^{(\mathrm{tor})}$
have the form,
\begin{equation}\label{eq:rhopm}
  \rho=\left[
    \frac{\kappa-1}{\kappa}
    \frac{W_{\mathrm{in}}-W}{K+K_{m}\mathcal{L}^{\kappa-1}}
  \right]^{\frac{1}{\kappa-1}},
  \quad
  p_{m}^{(\mathrm{tor})}=K_{m}\mathcal{L}^{\kappa-1}
  \left[
    \frac{\kappa-1}{\kappa}
    \frac{W_{\mathrm{in}}-W}{K+K_{m}\mathcal{L}^{\kappa-1}}
  \right]^{\frac{\kappa}{\kappa-1}},
\end{equation}
where $K$, $K_{m}$, and $\kappa$ are the constants in the equations
of state, $p=Kw^{\kappa}$ and $p_{m}^{(\mathrm{tor})}=K_{m}\mathcal{L}^{\kappa-1}w^{\kappa}$.
Here, $p$ is the plasma pressure and $w$ is the specific enthalpy.
Following ref.~\cite{Kom06}, we take that $\kappa=4/3$. The form of the
disk depends on the potential $W$,
\begin{equation}\label{eq:genfunW}
  W(r,\theta)=\frac{1}{2}\ln
  \left|
    \frac{\mathcal{L}}{\mathcal{A}}
  \right|.
\end{equation}
The parameter $W_{\mathrm{in}}$ in eq.~(\ref{eq:rhopm}) is the
value of $W$ at the border of the disk.

Equations~(\ref{eq:UB})-(\ref{eq:genfunW}) completely define all
the characteristics of the disk. First, using eq.~(\ref{eq:genfunW}),
we get that points $(r,\theta)$ inside the disk obey the condition
$W\leq W_{\mathrm{in}}$. Then, we apply eq.~(\ref{eq:rhopm}) to
find $\rho$ and $p_m^{(\mathrm{tor})}$. We define the effective toroidal field
$|\mathbf{B}|^{(\mathrm{tor})}=\sqrt{2p_{m}^{(\mathrm{tor})}}$. The
maximal value of $|\mathbf{B}|^{(\mathrm{tor})}$ is equated to the
strength expected in the disk. It gives us one of the equations to
define the constants $K$ and $K_{m}$. Another equation appears if
we associate $\rho_{\mathrm{max}}$ with maximal plasma density present
in the disk. Finally, eq.~(\ref{eq:UB}) gives us the rest of the
parameters.

\begin{figure}
  \centering
  \subfigure[]
  {\label{fig:f3a}
  \includegraphics[scale=.35]{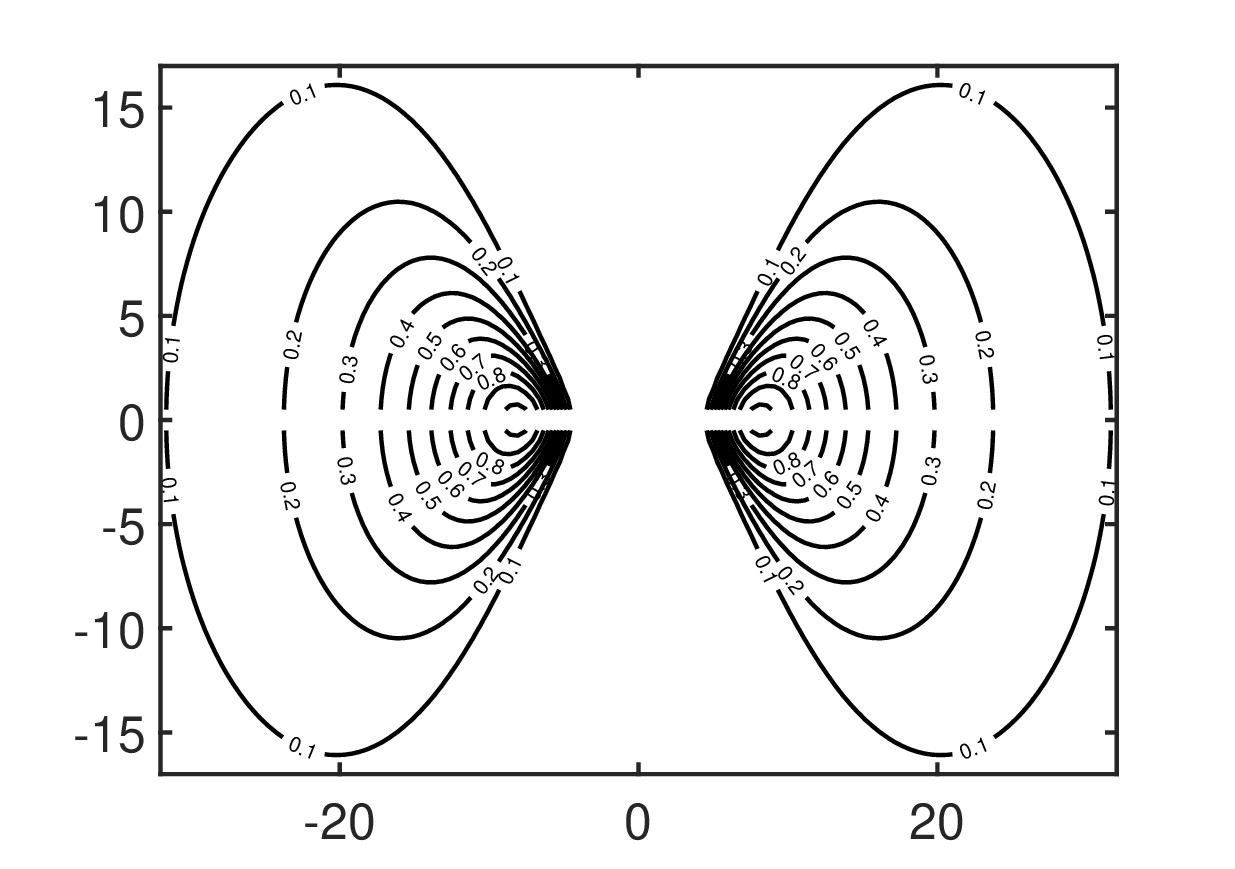}}
  \subfigure[]
  {\label{fig:f3b}
  \includegraphics[scale=.35]{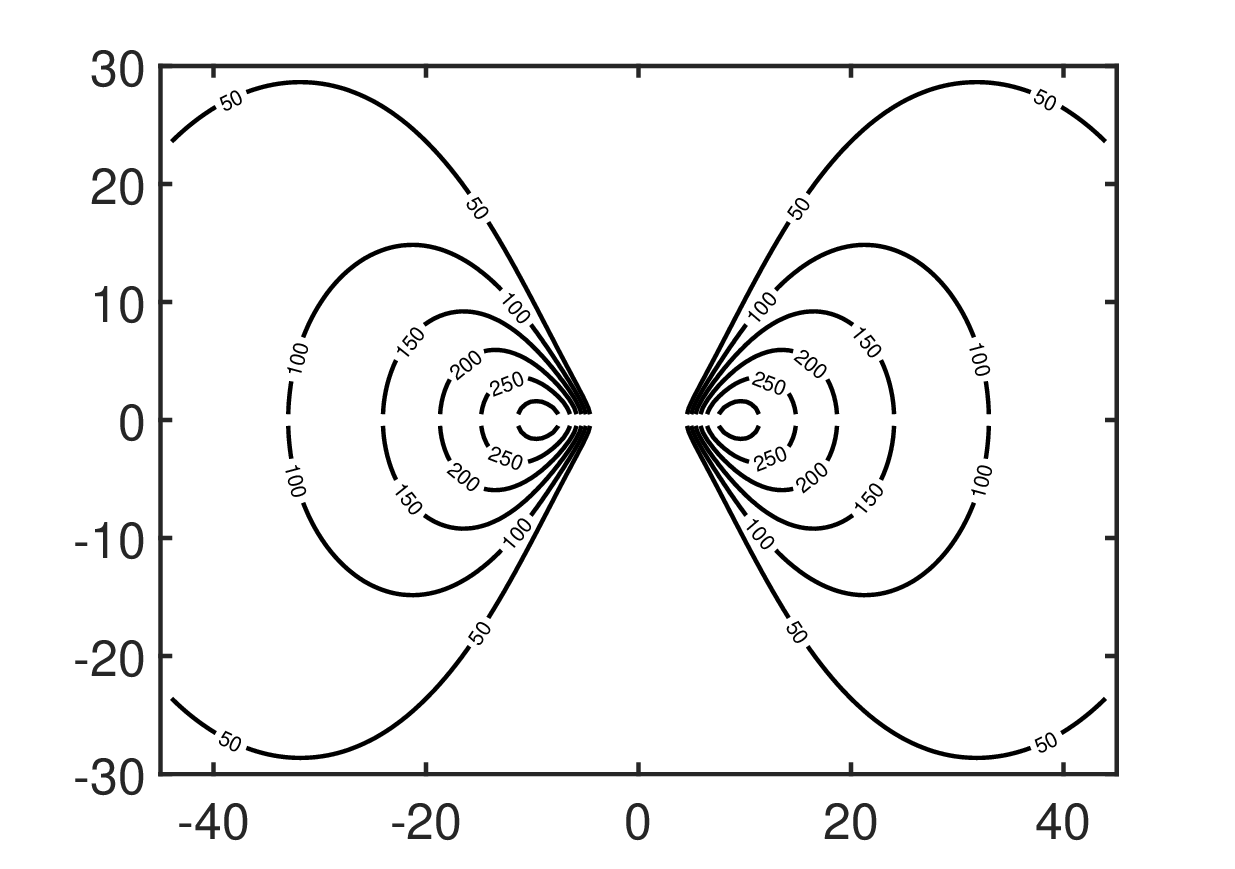}}
  \\
  \subfigure[]
  {\label{fig:f3c}
  \includegraphics[scale=.35]{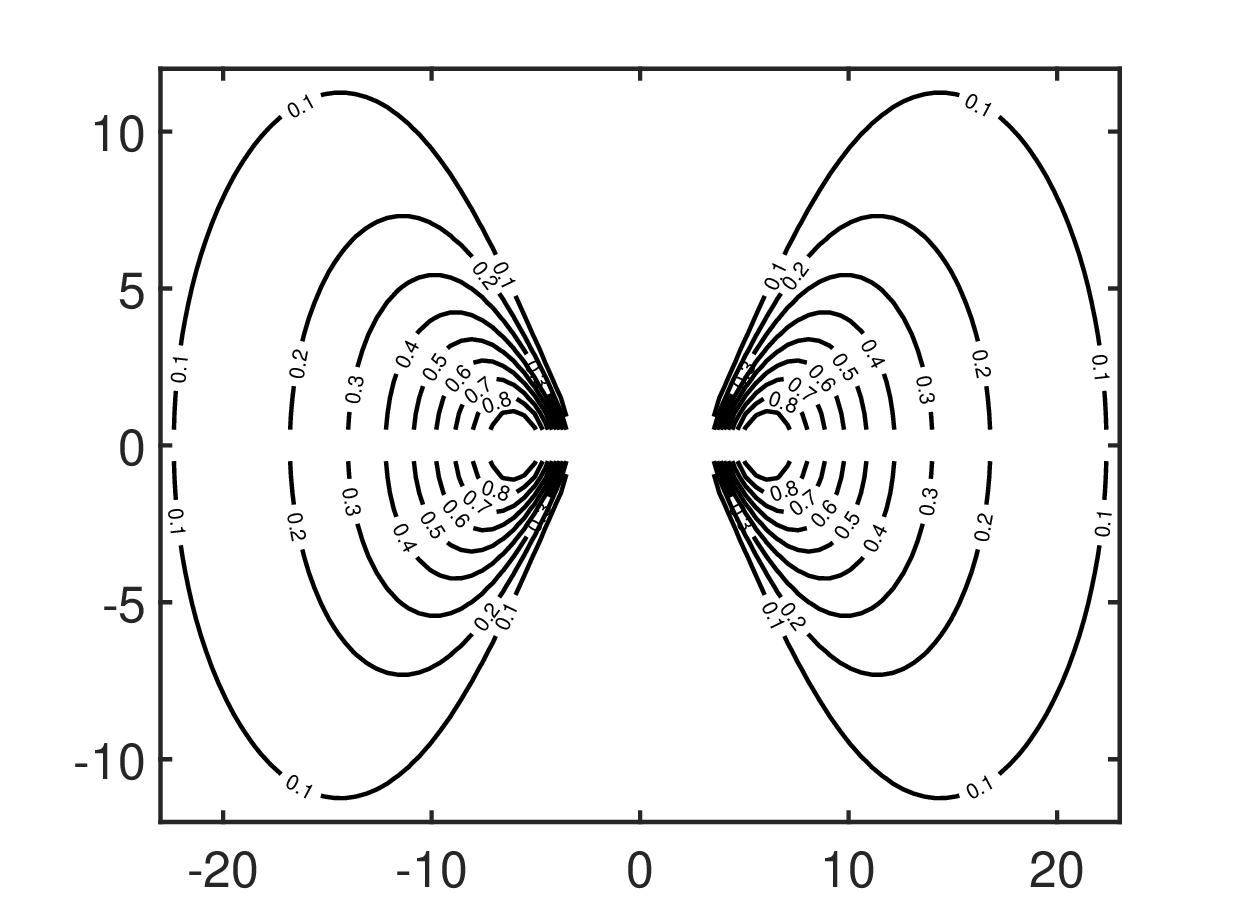}}
  \subfigure[]
  {\label{fig:f3d}
  \includegraphics[scale=.35]{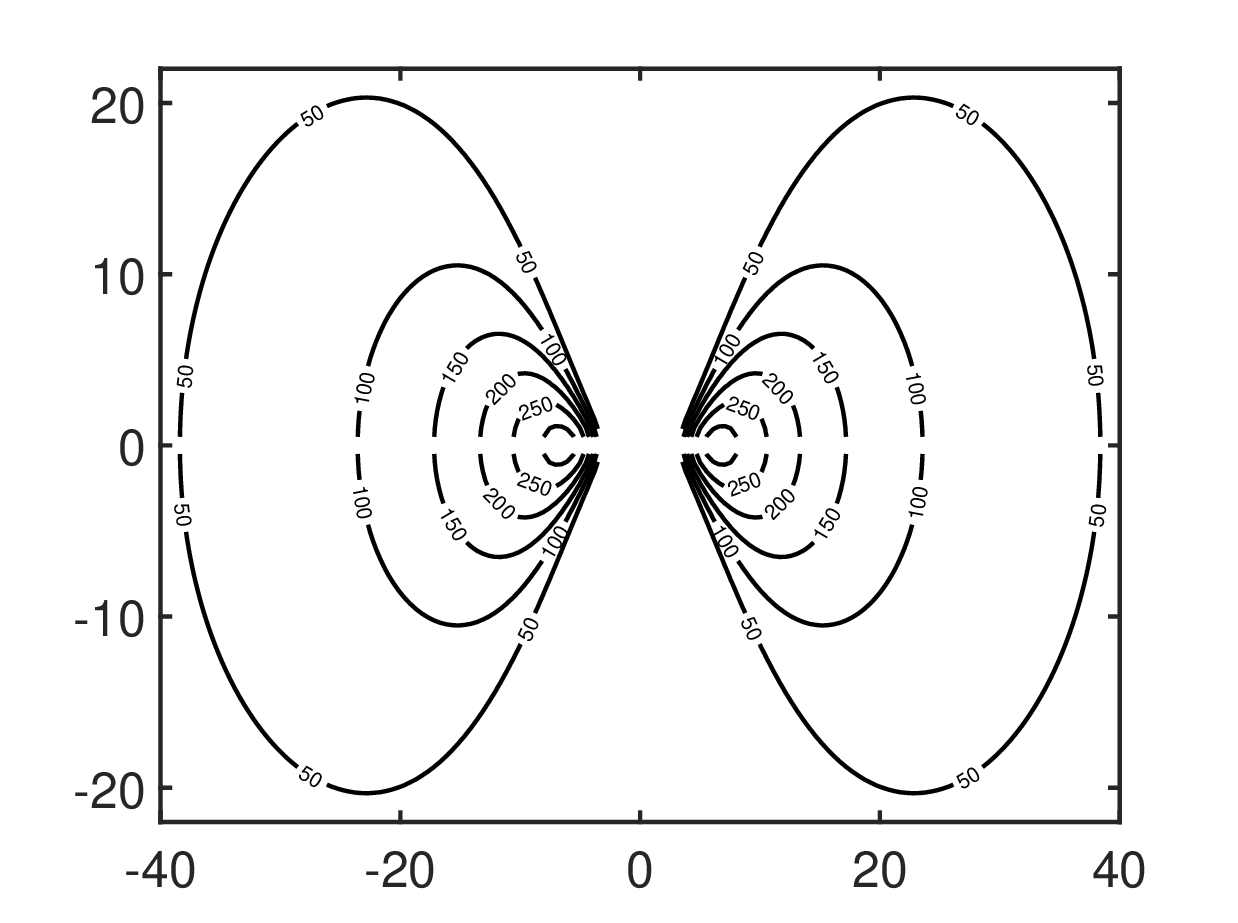}}
  \\
  \subfigure[]
  {\label{fig:f3e}
  \includegraphics[scale=.35]{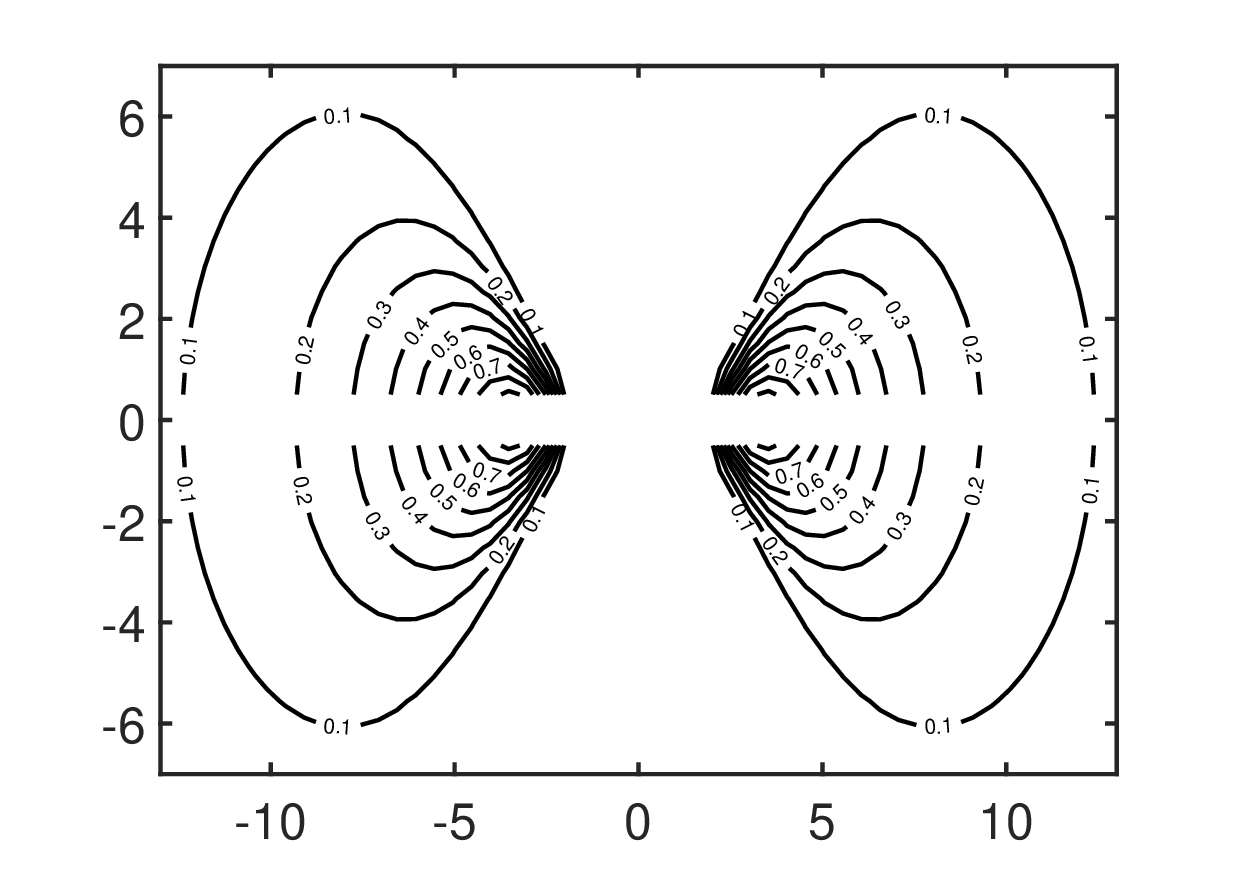}}
  \subfigure[]
  {\label{fig:f3f}
  \includegraphics[scale=.35]{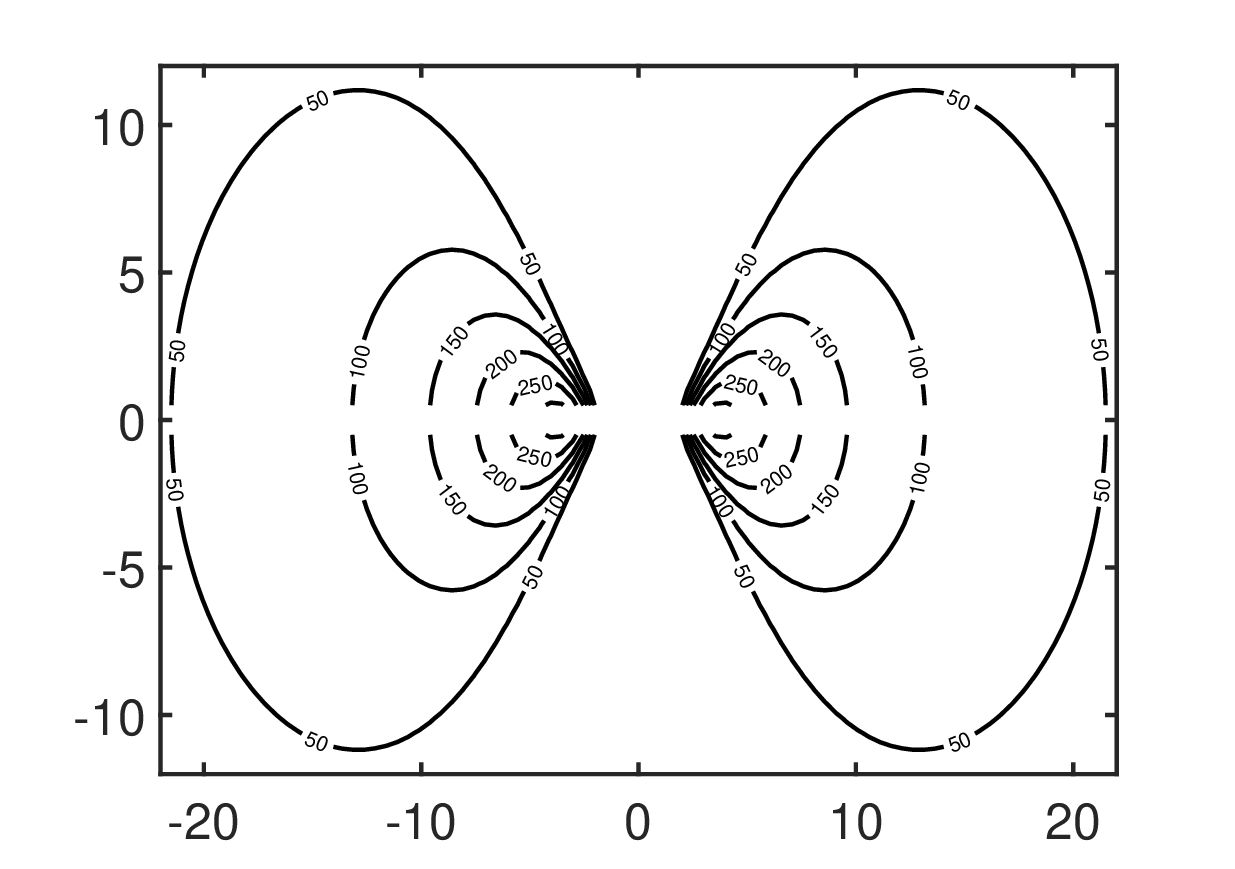}}
  \protect 
\caption{The distributions of the normalized electron number density $n_{e}/10^{18}\,\text{cm}^{-3}$
[panels~(a), (c), and~(e)] and the effective toroidal magnetic
field $|\mathbf{B}|^{(\mathrm{tor})}=\sqrt{2p_{m}^{(\mathrm{tor})}}$,
in Gauss, [panels~(b), (d), and~(f)] for different spins of
BH. Panels~(a) and~(b): $a=2\times10^{-2}M$ ($z=10^{-2}$), $\tilde{K}=2.55\times10^{-31}$,
and $\tilde{K}_{m}=3.59\times10^{-41}$; panels~(c) and~(d): $a=0.5M$
($z=0.25$) $\tilde{K}=3.6\times10^{-31}$, and $\tilde{K}_{m}=4.66\times10^{-41}$;
panels~(e) and~(f): $a=0.9M$ ($z=0.45$) $\tilde{K}=6.5\times10^{-31}$,
and $\tilde{K}_{m}=7.2\times10^{-41}$. The distances in the horizontal
and vertical axes are in $r_{g}$.\label{fig:nepmdistr}}
\end{figure}

The distributions of the normalized electron number density $n_{e}/10^{18}\,\text{cm}^{-3}$,
where $n_{e}=\rho/m_{p}$, and the effective toroidal magnetic field
$|\mathbf{B}|^{(\mathrm{tor})}=\sqrt{2p_{m}^{(\mathrm{tor})}}$, measured
in Gauss, are shown in figure~\ref{fig:nepmdistr}. for different spins
of BH. We use the dimensionless variables $\tilde{K}=r_{g}^{4(1-\kappa)}K$
and $\tilde{K}_{m}=r_{g}^{2(1-\kappa)}K_{m}$. In all cases in figure~\ref{fig:nepmdistr},
$n_{e}^{(\mathrm{max})}=10^{18}\,\text{cm}^{-3}$ and $|\mathbf{B}|_{\mathrm{max}}^{(\mathrm{tor})}=320\,\text{G}$.
Figure~\ref{fig:nepmdistr} corresponds to $W_{\mathrm{in}} = - 10^{-5}$ and $\lambda_{0}=0.6(\lambda_{\mathrm{mb}}+\lambda_{\mathrm{ms}})$,
where $\lambda_{\mathrm{mb,ms}}=\lambda(x_{\mathrm{mb,ms}})$ and
\begin{equation}
  \lambda(x)=\frac{x^{2}-z\sqrt{2x}+z^{2}}{\sqrt{2}x^{3/2}-\sqrt{2x}+z}.
\end{equation}
That is, we study the disk corotating with BH. The quantities $x_{\mathrm{mb}}$
and $x_{\mathrm{ms}}$ are the radii of the marginally bound and marginally
stable Keplerian orbits~\cite{BarPreTeu72}.

The model in ref.~\cite{Kom06} provides only the toroidal magnetic field.
It was proved in refs.~\cite{Tay73,MarTay73} that only a toroidal or only a poloidal
magnetic field are unstable. The superposition of the these fields
can be stable since it has a nonzero linking number and, thus, a nonzero
magnetic helicity. That is why we suppose that a nonzero poloidal
field is present in the disk. We consider two models for such a field.

First, we take the following vector potential:
\begin{equation}\label{eq:Atphi}
  A_{t}=Ba
  \left[
    1-\frac{rr_{g}}{2\Sigma}(1+\cos^{2}\theta)
  \right],
  \quad
  A_{\phi}=-\frac{B}{2}
  \left[
    r^{2}+a^{2}-\frac{a^{2}rr_{g}}{\Sigma}(1+\cos^{2}\theta)
  \right]
  \sin^{2}\theta,
\end{equation}
where $\Sigma$ is given in eq.~(\ref{eq:ingmet}). For the first time, $A_{\mu}$
in eq.~(\ref{eq:Atphi}) was proposed in ref.~\cite{Wal74} to describe
the electromagnetic field in the vicinity of a rotating BH which asymptotically
equals to a constant and uniform magnetic field $\mathbf{B}=B\mathbf{e}_{z}$.
Analogous magnetic field configuration is used, e.g., in ref.~\cite{NerSemTka09}
to explain the cosmic rays acceleration by BH.

The assumption of the constant $B$ in eq.~(\ref{eq:Atphi}) is unphysical
since the magnetic field should vanish towards the outer edge of the
disk. That is, following ref.~\cite{BlaPay82}, we assume that $B\propto B_{0}r^{-5/4}$.
The strength $B_{0}$ at $r\sim r_{g}$ is chosen to be close to $|\mathbf{B}|_{\mathrm{max}}^{(\mathrm{tor})}$.
Previously, such a model of the
poloidal field was used in refs.~\cite{Dvo23a,Dvo23b,Dvo23c} in the whole space outside BH. Now, we suppose that
it exists only inside the disk given by the condition $W\leq W_{\mathrm{in}}$.

Second, we use the poloidal field, proposed in ref.~\cite{FraMei09},
\begin{equation}\label{eq:Aphi}
  A_{\phi}=b\rho,
\end{equation}
where $b$ is a constant parameter and $\rho$ in given in eq.~(\ref{eq:rhopm}).
It should be noted that the poloidal field, corresponding to eq.~(\ref{eq:Aphi}),
exists only inside the disk defined by $W\leq W_{\mathrm{in}}$.

The only nonzero components of $F_{\mu\nu}$, corresponding to eq.~(\ref{eq:Aphi}),
are $F_{r\phi}=b\partial_{r}\rho$ and $F_{\theta\phi}=b\partial_{\theta}\rho$.
Using eq.~(\ref{eq:Fmunudisk}), we find the magnetic pressure $p_{m}^{(\mathrm{pol})}=-g_{\mu\nu}B^{\mu}B^{\nu}/2$,
\begin{equation}\label{eq:pmpol}
  p_{m}^{(\mathrm{pol})}=\frac{b^{2}}{2\sin^{2}\theta\Sigma}
  \left|
    \frac{\mathcal{L}}{\mathcal{A}}
  \right|
  \left[
    \frac{1}{\Delta}(\partial_{\theta}\rho)^{2}+(\partial_{r}\rho)^{2}
  \right],
\end{equation}
where $\Delta$ is given in eq.~(\ref{eq:ingmet}). We introduce
the effective poloidal magnetic field $|\mathbf{B}|^{(\mathrm{pol})}=\sqrt{2p_{m}^{(\mathrm{pol})}}$.
The parameter $b$ in eq.~(\ref{eq:Aphi}) is fixed when we suppose
that the maximal value of $|\mathbf{B}|^{(\mathrm{pol})}$ equals
to $|\mathbf{B}|_{\mathrm{max}}^{(\mathrm{tor})}$, which is defined
earlier.

\begin{figure}
  \centering
  \subfigure[]
  {\label{fig:f4a}
  \includegraphics[scale=.35]{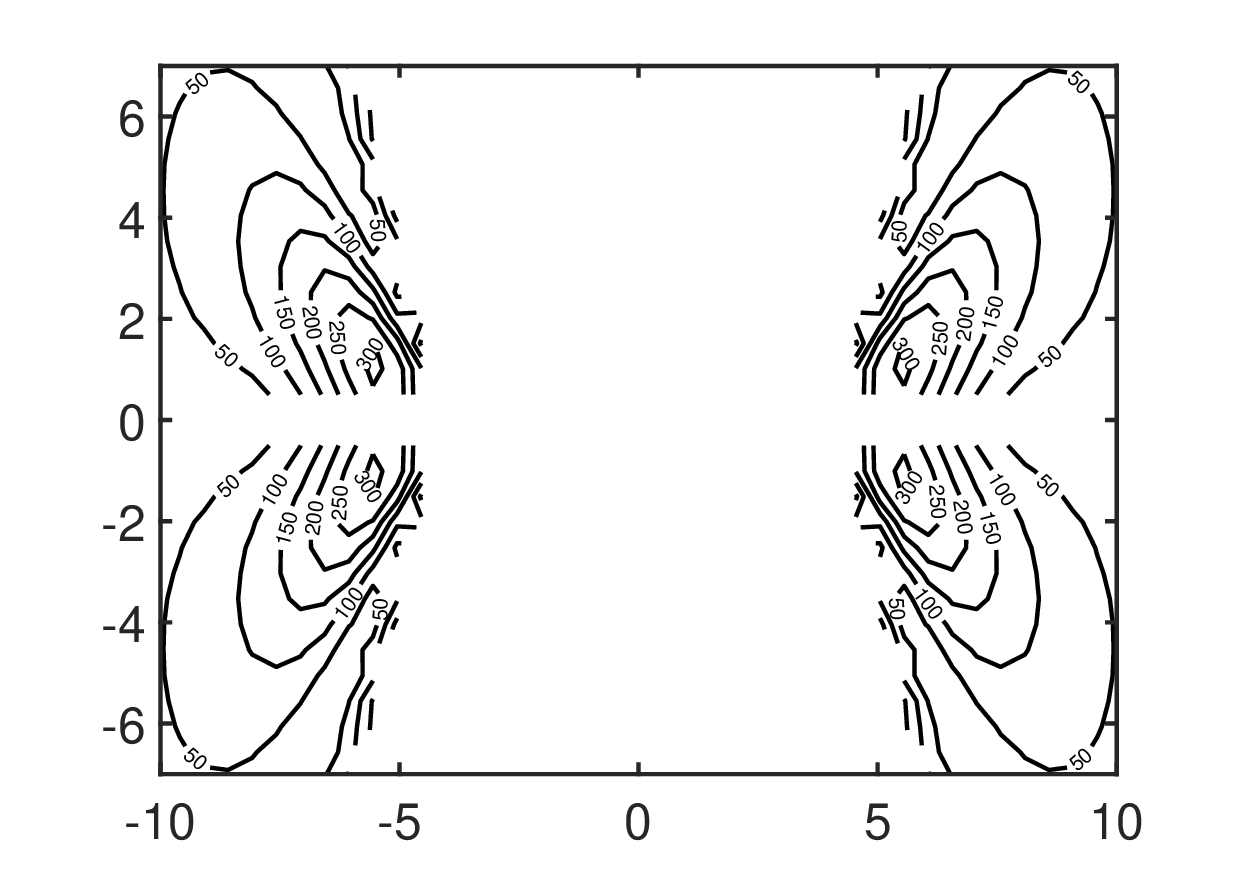}}
  \subfigure[]
  {\label{fig:f4b}
  \includegraphics[scale=.33]{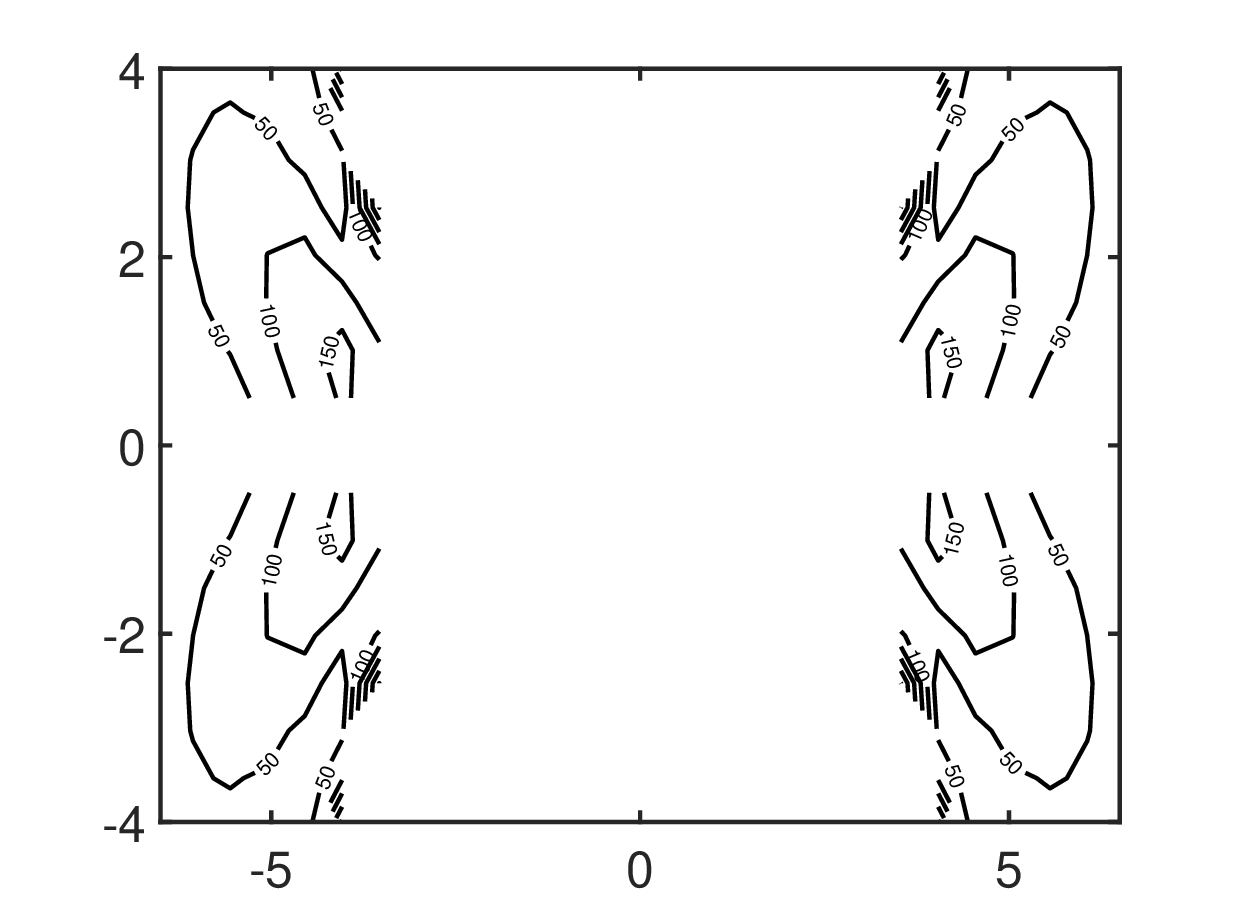}}
  \protect 
\caption{The distribution of the effective poloidal magnetic field $|\mathbf{B}|^{(\mathrm{pol})}=\sqrt{2p_{m}^{(\mathrm{pol})}}$,
in Gauss, corresponding to eq.~(\ref{eq:Aphi}), for different spins
of BH. Panel~(a): $a=2\times10^{-2}M$ ($z=10^{-2}$) and $\tilde{b}=6.42\times10^{-48}$;
panel~(b): $a=0.5M$ ($z=0.25$) and $\tilde{b}=1.78\times10^{-48}$.
The rest of the parameters is the same as in figure~\ref{fig:nepmdistr}.\label{fig:pmplosoph}}
\end{figure}

We show the distribution of $|\mathbf{B}|^{(\mathrm{pol})}$ for eq.~(\ref{eq:pmpol})
for different spins of BH in figure~\ref{fig:pmplosoph}. We present the
situations when $z=10^{-2}$ and $z=0.5$. In the case of a rapidly
rotating BH with $z=0.9$, $|\mathbf{B}|^{(\mathrm{pol})}$ has a
very sharp maximum, which is hardly visible in a contour plot. Therefore,
we omit it.

\section{Solution of the Schr\"{o}dinger equation for neutrinos below the equatorial
plane\label{sec:BELOW}}

In our problem, the flux of incoming neutrinos is both above and below
the equatorial plane. We solve eq.~(\ref{eq:Schreq}) only for up
particles. The solution for down particles can be reconstructed automatically
applying the symmetry reasons.

Suppose that the effective Hamiltonian in eq.~(\ref{eq:Schreq})
for up particles is $\hat{H}_{u}$. The Hamiltonian for down particles
is $\hat{H}_{d}=-\hat{H}_{u}^{*}$, where the star means the complex
conjugation. The formal solution of eq.~(\ref{eq:Schreq}) is
\begin{equation}\label{eq:psiu}
  \psi_{u}(x)=
  \left[
    1-\mathrm{i}\int_{-\infty}^{x}\hat{H}_{u}(x')\mathrm{d}x'-
    \frac{1}{2}\int_{-\infty}^{x}\hat{H}_{u}(x')\mathrm{d}x'
    \int_{-\infty}^{x'}\hat{H}_{u}(x'')\mathrm{d}x''+\dotsb
  \right]
  \psi_{-\infty},
\end{equation}
where $\psi_{-\infty}^{\mathrm{T}}=(1,0)$ is the initial condition.
Taking the complex conjugation of eq.~(\ref{eq:psiu}), we get
\begin{align}\label{eq:psiucc}
  \psi_{u}^{*}(x)= &
  \left[
    1+\mathrm{i}\int_{-\infty}^{x}\hat{H}_{u}^{*}(x')\mathrm{d}x'-
    \frac{1}{2}\int_{-\infty}^{x}\hat{H}_{u}^{*}(x')\mathrm{d}x'
    \int_{-\infty}^{x'}\hat{H}_{u}^{*}(x'')\mathrm{d}x''+\dotsb
  \right]\psi_{-\infty}
  \nonumber
  \\
  & =
  \left[
    1-\mathrm{i}\int_{-\infty}^{x}\hat{H}_{d}(x')\mathrm{d}x'-
    \frac{1}{2}\int_{-\infty}^{x}\hat{H}_{d}(x')\mathrm{d}x'
    \int_{-\infty}^{x'}\hat{H}_{d}(x'')\mathrm{d}x''+\dotsb
  \right]
  \psi_{-\infty}.
\end{align}
Thus, $\psi_{d}(x)=\psi_{u}^{*}(x)$ since the initial condition is
real and coincide for both particles.

Therefore $P_{\mathrm{LL}}^{(u)}=|\psi_{+\infty}^{(u,\mathrm{L})}|^{2}=P_{\mathrm{LL}}^{(d)}=|\psi_{+\infty}^{(d,\mathrm{L})}|^{2}$.
When we map the flux of outgoing down particles, we should take into
account that $\phi_{\mathrm{obs}}^{(d)}=\phi_{\mathrm{obs}}^{(u)}$
and $\theta_{\mathrm{obs}}^{(d)}=\pi-\theta_{\mathrm{obs}}^{(u)}$.

\end{document}